\documentclass[sigconf]{acmart}
\usepackage{multirow}
\usepackage{graphicx}
\usepackage{subfigure}
\usepackage{soul} 
\usepackage{color, xcolor} 
\AtBeginDocument{%
  \providecommand\BibTeX{{%
    \normalfont B\kern-0.5em{\scshape i\kern-0.25em b}\kern-0.8em\TeX}}}

\copyrightyear{2023} 
\acmYear{2023} 
\setcopyright{acmlicensed}\acmConference[MM '23]{Proceedings of the 31st
ACM International Conference on Multimedia}{October 29-November 3,
2023}{Ottawa, ON, Canada}
\acmBooktitle{Proceedings of the 31st ACM International Conference on
Multimedia (MM '23), October 29-November 3, 2023, Ottawa, ON, Canada}
\acmPrice{15.00}
\acmDOI{10.1145/3581783.3613767}
\acmISBN{979-8-4007-0108-5/23/10}

\acmSubmissionID{1207}

\begin{document}

\title{UMMAFormer: A Universal Multimodal-adaptive Transformer
Framework for Temporal Forgery Localization}





\author{Rui Zhang}
\email{zhangrui1997@stu.scu.edu.cn}
\orcid{0000-0002-6889-1247}
\affiliation{%
  \institution{School of Cyber Science and Engineering, Sichuan University}
  \streetaddress{Chuanda Road}
  \city{Chengdu}
  \state{Sichuan}
  \country{China}
  \postcode{610065}
}

\author{Hongxia Wang}
\authornote{Corresponding author.}
\email{hxwang@scu.edu.cn}
\orcid{0000-0001-6294-3272}
\affiliation{%
  \institution{School of Cyber Science and Engineering, Sichuan University}
  \streetaddress{Chuanda Road}
  \city{Chengdu}
  \state{Sichuan}
  \country{China}
  \postcode{610065}
}

\author{Mingshan Du}
\email{2022226245054@stu.scu.edu.cn}
\orcid{0009-0008-7820-144X}
\affiliation{%
  \institution{School of Cyber Science and Engineering, Sichuan University}
  \streetaddress{Chuanda Road}
  \city{Chengdu}
  \state{Sichuan}
  \country{China}
  \postcode{610065}
}

\author{Hanqing Liu}
\email{liuhanqing0520@stu.scu.edu.cn}
\orcid{0000-0002-1295-4851}
\affiliation{%
  \institution{School of Cyber Science and Engineering, Sichuan University}
  \streetaddress{Chuanda Road}
  \city{Chengdu}
  \state{Sichuan}
  \country{China}
  \postcode{610065}
}

\author{Yang Zhou}
\email{yzhoulv@stu.scu.edu.cn}
\orcid{0000-0002-9148-4316}
\affiliation{%
  \institution{School of Cyber Science and Engineering, Sichuan University}
  \streetaddress{Chuanda Road}
  \city{Chengdu}
  \state{Sichuan}
  \country{China}
  \postcode{610065}
}

\author{Qiang Zeng}
\email{zengqiang@stu.scu.edu.cn}
\orcid{0009-0002-6938-4877}
\affiliation{%
  \institution{School of Cyber Science and Engineering, Sichuan University}
  \streetaddress{Chuanda Road}
  \city{Chengdu}
  \state{Sichuan}
  \country{China}
  \postcode{610065}
}

\renewcommand{\shortauthors}{Rui Zhang et al.}

\begin{abstract}

The emergence of artificial intelligence-generated content (AIGC) has raised concerns about the authenticity of multimedia content in various fields. However, existing research for forgery content detection has focused mainly on binary classification tasks of complete videos, which has limited applicability in industrial settings. To address this gap, we propose UMMAFormer, a novel universal transformer framework for temporal forgery localization (TFL) that predicts forgery segments with multimodal adaptation. Our approach introduces a Temporal Feature Abnormal Attention (TFAA) module based on temporal feature reconstruction to enhance the detection of temporal differences. We also design a Parallel Cross-Attention Feature Pyramid Network (PCA-FPN) to optimize the Feature Pyramid Network (FPN) for subtle feature enhancement. To evaluate the proposed method, we contribute a novel Temporal Video Inpainting Localization (TVIL) dataset specifically tailored for video inpainting scenes. Our experiments show that our approach achieves state-of-the-art performance on benchmark datasets, including Lav-DF, TVIL, and Psynd, significantly outperforming previous methods.  The code and data are available at https://github.com/ymhzyj/UMMAFormer/.

\end{abstract}

\ccsdesc[500]{Computing methodologies~Computer vision}
\ccsdesc[500]{Applied computing~Investigation techniques}
\keywords{temporal forgery localization, transformer, multimodal-adaptive}


\maketitle

\section{Introduction}

The rapid development of advanced multimedia editing software enabled by artificial intelligence-generated content (AIGC)~\cite{DBLP:journals/corr/abs-2209-00796, DBLP:journals/csur/AldausariSMM23,DBLP:journals/corr/abs-2005-05535, DBLP:conf/mm/LiH0SYY22, DBLP:conf/mm/HegdePMNJ22,DBLP:conf/mm/WangZ22,DBLP:conf/mm/CaiLTYT22, DBLP:conf/cvpr/0031LQGC22} has raised concerns about its potential misuse, such as manipulating public opinion and fabricating evidence. This has led to a growing interest in developing methods for detecting manipulated content in multimedia forensics, with a primary focus on deepfake detection~\cite{DBLP:conf/mm/ZhangLHW0G22, DBLP:conf/cvpr/ZhaoZ0WZY21, DBLP:conf/mm/KwakCYLHO22,DBLP:conf/mm/DengMY0D22, DBLP:conf/iccv/ZhouL21} in facial and audio media. Despite the promising results demonstrated by these methods in a variety of benchmarks~\cite{DBLP:conf/nips/KhalidTKW21,DBLP:journals/corr/abs-1910-08854,DBLP:conf/mm/ZiCCMJ20,DBLP:conf/iccv/RosslerCVRTN19,DBLP:conf/nips/FrankS21, DBLP:journals/corr/abs-2006-07397}, their mainstream adoption by the industry remains limited due to the constraint of binary classification tasks.  These methods are inadequate for identifying the temporal boundaries of manipulations, which is crucial for practical applications. Further research is necessary to develop techniques that can accurately locate temporal boundaries of manipulations in multimedia content and promote the responsible use of AIGC for the betterment of society.

\begin{figure*}[htbp]
  \includegraphics[width=2\columnwidth]{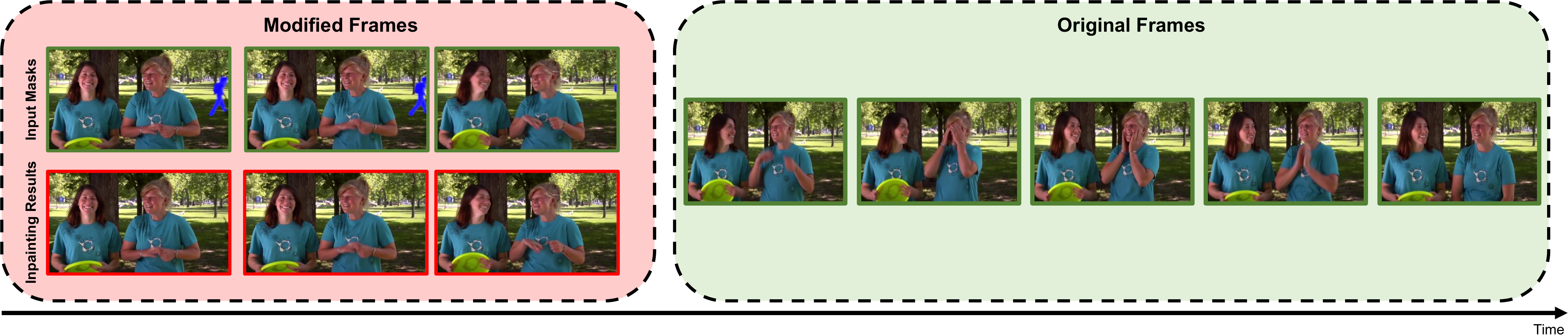}
  \centering
  \caption{We show a collection of keyframes extracted from manipulated videos where a person has been removed, posing a serious threat to digital evidence integrity. The original person mask is displayed on the left with the corresponding video inpainting results underneath it, while the unmanipulated results are shown on the right with a green background. Of the 144 frames in the video, only 11 involve person removal, and it only takes minor modifications to a small section of the video to achieve this. This manipulation technique has drawn significant attention in forensic video analysis because manipulated videos may be presented as genuine evidence in legal proceedings and can be difficult to detect using classification-based methods.}
  \Description{We presented a collection of keyframes extracted from videos that have been attacked by person removal, which poses a serious threat to the integrity of digital evidence.}
  \label{fig:videosample}
\end{figure*}

Recent studies~\cite{DBLP:conf/mm/ChughGDS20,DBLP:journals/corr/abs-2204-06228, DBLP:conf/icpr/ZhangS22,DBLP:conf/cvpr/HeGCZYSSS021} have proposed a new task called temporal forgery localization (TFL) to overcome the limitations of binary classification in detecting manipulated content in multimedia. TFL aims to locate the start and end timestamps of manipulated segments, providing a wider range of application scenarios and helping users better understand the results of forgery detection. TFL is similar to temporal action localization (TAL)~\cite{DBLP:conf/cvpr/HeilbronEGN15,DBLP:journals/pami/VahdaniT23,DBLP:conf/mm/RenXSYL21} and follows a similar process: pre-processing the video or audio data using a pre-trained feature extractor, enhancing the representation of feature vectors with a designed neural network architecture, and decoding the feature vectors using regression and classification heads to obtain the start and end times of each action segment and their corresponding categories. Note that this process may vary depending on specific task requirements.

TFL tasks present unique challenges compared to TAL. Firstly, real-world scenarios often involve various modalities, including audio-only, visual-only, and audio-visual data, requiring separate models for manipulation detection and potentially delaying TFL technology development. Secondly, unmodified or real samples are essential in TFL, just like background samples, but they are often neglected in TAL. Thirdly, manipulation changes are usually more subtle than action changes, with minor alterations like a single word or short pronunciation time making detection more challenging. Finally, the lack of available datasets is a significant bottleneck for TFL. Most multimedia forgery datasets evaluate manipulation performance over the entire video or audio. Only a few studies have validated TFL performance, limited to a single dataset such as Lav-DF~\cite{DBLP:journals/corr/abs-2204-06228} for the visual domain and Psynd~\cite{DBLP:conf/icpr/ZhangS22} for the audio domain. Besides, available datasets for TFL~\cite{DBLP:journals/corr/abs-2204-06228,DBLP:conf/icpr/ZhangS22,DBLP:conf/cvpr/HeGCZYSSS021} and deepfake detection~\cite{DBLP:conf/nips/FrankS21,DBLP:conf/nips/KhalidTKW21,DBLP:conf/mm/ZiCCMJ20} primarily focus on facial manipulations and speech forgeries, while AIGC still poses a threat in other scenarios. This narrow scope limits the potential applications for forgery detection and TFL. For instance, video inpainting~\cite{DBLP:conf/cvpr/0031LQGC22} techniques can remove specific objects from videos, leading to fabricated evidence, as shown in Figure~\ref{fig:videosample}. Based on the above observations, we propose the following work.

For different modalities of multimedia, we propose a novel universal multimodal-adaptive transformer framework for TFL called UMMAFormer. The framework aims to predict forgery segments and their corresponding start and end timestamps in untrimmed videos or audios. Transformer-based models~\cite{DBLP:conf/nips/VaswaniSPUJGKP17,DBLP:conf/mm/GuoTDDK22,DBLP:conf/iccv/HuS21,DBLP:journals/taslp/LiLZFZ21} have demonstrated excellent performance in various tasks and can adapt to different modality feature inputs. Therefore, we build a universal multimodal adaptive framework based on the transformer block that can be used for TFL tasks involving different modalities of data.

In order to fully utilize real samples, we design a Temporal Feature Abnormal Attention (TFAA) module based on temporal feature reconstruction. Our motivation is based on the observation that compared to TAL which relies on spatial content to recognize specific types of actions, TFL relies more on temporal features that reflect the changes caused by spatial content manipulations. The underlying difference in feature distribution between manipulated and real segments can be considered a universal feature of multimedia manipulation that exists across any modality of input. By incorporating TFAA, our method enhances the detection of temporal differences, leading to improved TFL performance across different input feature modalities.

For analyzing short video clips with subtle variations, Feature Pyramid Network (FPN)~\cite{DBLP:conf/cvpr/LinDGHHB17} is a commonly used solution that effectively enhances subtle features. We further optimize FPN by introducing a parallel structure and proposing a Parallel Cross-Attention Feature Pyramid Network (PCA-FPN). PCA-FPN significantly improves the performance of small manipulated segments localization.

To advance research further, it is critical to create a new dataset for a novel scenario and providing new evaluation benchmarks for advancing research in TFL tasks. We introduce a novel temporal video inpainting localization dataset called TVIL for training and evaluation of TFL tasks. As per our knowledge, we are the first ones to present a TFL dataset that is tailored for video inpainting scenes. Our dataset is built on the YouTube-VOS 2018~\cite{DBLP:conf/eccv/XuYFYYLPCH18} dataset. We employ XMEM~\cite{DBLP:conf/eccv/ChengS22} to annotate segmentation masks for all frames in the dataset, and then use four different video inpainting models~\cite{DBLP:conf/cvpr/0031LQGC22,DBLP:conf/eccv/ZhangFL22,DBLP:conf/iccv/0019DHSLS0D021,DBLP:conf/eccv/ZengFC20} to erase objects in random time periods. We acquire 4453 tampered videos with annotations, divided into training, validation, and testing sets according to the same proportions as the original dataset. 

We conduct extensive experiments on three benchmark datasets, Lav-DF~\cite{DBLP:journals/corr/abs-2204-06228}, Psynd~\cite{DBLP:conf/icpr/ZhangS22}, and TVIL to evaluate the effectiveness of our proposed method. The results demonstrate that our approach achieves state-of-the-art performance on these datasets, outperforming the previous best results by a significant margin.

In summary, our contributions are:
\begin{itemize}
  \item  We introduce UMMAFormer, a novel universal transformer framework for multimedia temporal forgery localization that can be applied to various modalities of input. 
  \item We propose a TFAA module that enables the model to focus on temporal anomalies caused by spatial content tampering.
  \item We design PCA-FPN, a parallel cross-attention feature pyramid network, to improve the recognition and localization of ultrashort forgery segments.
  \item We present TVIL, a novel temporal video inpainting localization dataset, for research on TFL.
\end{itemize}

\section{RELATED WORK}

\subsection{Image-Level Forgery Detection}
Detecting manipulated content, especially deepfake~\cite{DBLP:conf/mm/LiH0SYY22, DBLP:journals/corr/abs-2005-05535}, has become a critical task in multimedia forensics. Significant efforts have been made to enhance image-level face forgery classification~\cite{DBLP:conf/cvpr/Cao0YCDY22, DBLP:conf/eccv/QianYSCS20, DBLP:conf/cvpr/LiBZYCWG20}. Early studies~\cite{DBLP:conf/icassp/NguyenYE19, DBLP:conf/iccv/RosslerCVRTN19} primarily relied on basic binary classifiers built upon existing backbone networks, suitable only for detecting low-quality generated images. With advancements in deepfake techniques, several approaches have been proposed to capture specific forgery traces. These approaches explore various features, including noise features~\cite{DBLP:conf/cvpr/ZhouHMD17, DBLP:conf/aaai/Wang_Chow_2023}, local texture characteristics~\cite{DBLP:conf/cvpr/ZhaoZ0WZY21, DBLP:conf/cvpr/Cao0YCDY22}, and frequency domain anomalies~\cite{DBLP:conf/mm/KwakCYLHO22, DBLP:conf/eccv/QianYSCS20}, to enhance detection capabilities. Unfortunately, these approaches overlooked the inclusion of temporal-level features, resulting in inconsistencies in discriminations for consecutive video frames due to variations in lighting, environmental factors, and other disturbances. As a consequence, they struggle to accurately differentiate genuine videos from forgeries and fail to identify temporal boundaries of the forgeries within the videos.

\subsection{Temporal-Level Forgery Detection}

Temporal-level forgery detection involves the classification of forgery at video or audio level and the TFL task, which is the main focus of this paper. The availability of various datasets~\cite{DBLP:conf/nips/KhalidTKW21,DBLP:journals/corr/abs-1910-08854,DBLP:conf/mm/ZiCCMJ20} has significantly contributed to the advancement of temporal-level forgery classification methods. Previous research has proposed different approaches to address this challenge. Hu et al.\cite{DBLP:journals/tcsv/HuLWQ22} presented a two-stream method, utilizing a temporal-level stream to extract temporal correlation features and analyze deepfake videos. Han et al.\cite{DBLP:journals/tbbis/HanHZLC21} introduced a two-stream network that uses temporal information and learnable spatial rich model (SRM) filters to detect fake videos at the video level. Song et al.\cite{DBLP:conf/mm/SongLFJCX22} utilized a symmetric transformer to enhance discrimination consistency between frames for video-level forgery classification. Additionally, Kwak et al.\cite{DBLP:conf/mm/KwakCYLHO22} developed a frequency feature masking method to classify real and fake audio in noisy environments. However, existing temporal-level forgery classification approaches usually treat temporal multimedia content as a cohesive entity, mainly focusing on distinguishing between real and manipulated content without verifying the authenticity of specific timestamps. To address this limitation and enhance the practical value of deepfake detection, the TFL task was introduced. Some studies~\cite{DBLP:conf/mm/ChughGDS20,DBLP:journals/corr/abs-2204-06228,DBLP:conf/icpr/ZhangS22,DBLP:conf/cvpr/HeGCZYSSS021} have focused on this task, but there is still room for significant improvement.

\subsection{Temporal Action Localization}
The goal of TAL is to localize the time intervals in a video when specific actions take place. Existing methods~\cite{DBLP:journals/pami/VahdaniT23} typically followed a general paradigm of feature extraction, feature enhancement, and prediction with post-processing. During the feature extraction stage, most TAL methods typically utilized pre-trained action recognition networks~\cite{DBLP:conf/eccv/WangXW0LTG16,DBLP:conf/iccv/Feichtenhofer0M19,DBLP:conf/icassp/HersheyCEGJMPPS17} to extract visual or audio-visual features. Given offline features, most algorithms mainly focus on enhancing features, by modeling action boundaries attention~\cite{DBLP:conf/aaai/ChenZWL22,DBLP:conf/cvpr/Lin0LWTWLHF21} and relationships~\cite{DBLP:conf/eccv/ZhangWL22,DBLP:journals/corr/abs-2101-08540,DBLP:conf/visapp/BagchiMFS22}. Some studies~\cite{DBLP:conf/iccv/LinLLDW19,DBLP:conf/iccv/TanT0W21,DBLP:conf/eccv/NagZSX22} also focused on proposing new regression and classification heads to further enhance the localization performance of the model.

\section{METHODOLOGY}
In this section, we introduce our universal multimodal-adaptive transformer framework, which aims to localizing temporal forgery in sequential multimedia data with various modalities. We have considered three scenarios, including visual-only, audio-only, or joint audio-visual modalities. Of course, the proposed approach can also be further extended to other types of tampered sequential multimedia data.

\begin{figure}[thbp]
	\centering
	\includegraphics[width=\columnwidth]{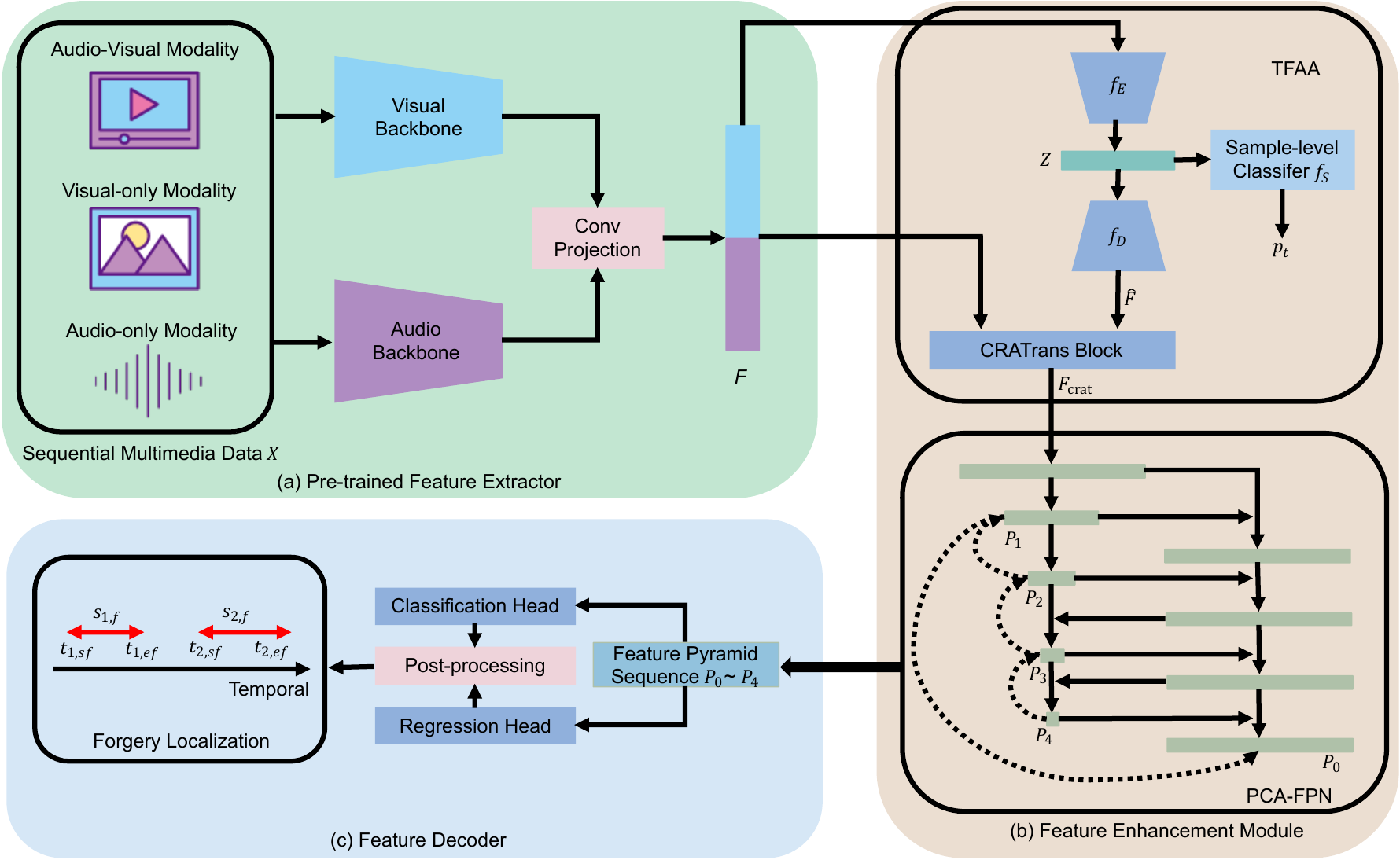}
	\caption{Illustration of the proposed UMMAFormer, which consists of three main components: (a) a pre-trained feature extractor that maps the sequential multimedia data $X$ to sequential features $F$, (b) a feature enhancement module that enhances the feature representation to multi-scale modified-sensitive features, and (c) a feature decoder that decodes the feature to localize forgeries in the data. }    
        \label{overview_frameworks}
\end{figure}

\subsection{Overview}
Our objective is to detect forgeries in untrimmed sequential multimedia data $X$ and locate the corresponding segments. Segments can be represented as $S=\left\{t_{n,sf},t_{n,ef}, s_{n,f}\right\}_{n=1}^{N_f}$, where $N_f$ is the number of detected modified segments, $t_{n,sf}$, $t_{n,ef}$, and $s_{n,f}$ are the start time, the end time, and the confidence score, respectively. To achieve this, $X$ is evenly split into $T$ segments $\left\{x_t\right\}^{T}_{t=1}$, and a feature sequence $F \in \mathbb{R}^{C \times T}$ is obtained using TSN~\cite{DBLP:conf/eccv/WangXW0LTG16} and BYOL-A~\cite{DBLP:conf/ijcnn/NiizumiTOHK21} as backbone networks for visual and audio data with concatenation of their features. Our proposed UMMAFormer framework, shown in Figure~\ref{overview_frameworks}, consists of a pre-trained feature extractor, a feature enhancement module based on a transformer-based network structure composed of the proposed TFAA module and PCA-FPN, and feature decoders for localization. We build on the ActionFormer~\cite{DBLP:conf/eccv/ZhangWL22} framework for our approach, with the feature decoding module directly utilized. Our proposed structure can also be extended to other TFL or TAL networks with similar processes.

\begin{figure}[htbp]
	\centering
	\includegraphics[width=\columnwidth]{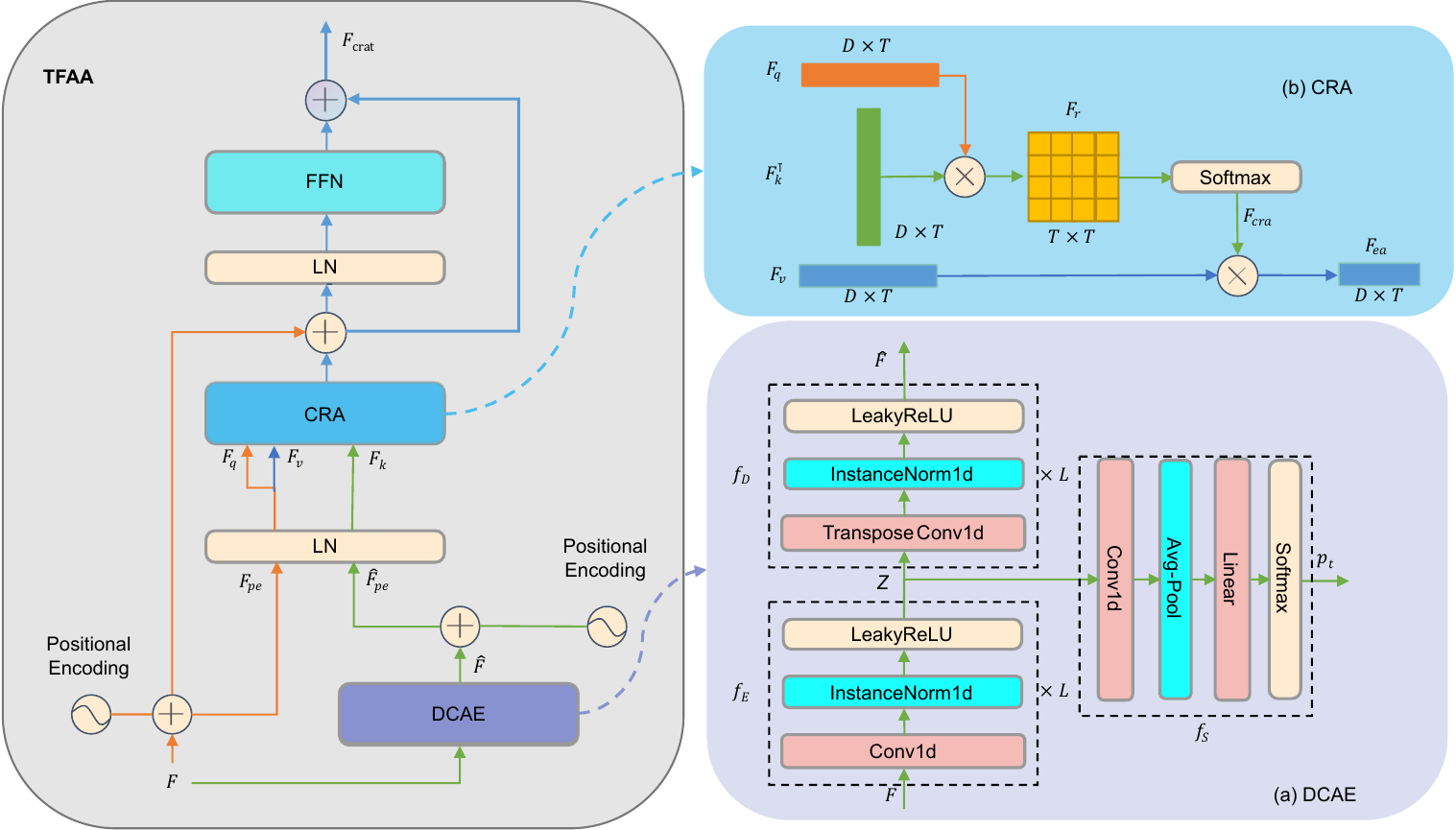}
	\caption{Illustration of the proposed TFAA module.}    
        \label{model_TFAA}
\end{figure}

\subsection{Temporal Feature Abnormal Attention}
\label{sec:TFAA}

To adapt to modified data from different modalities and make full use of real samples, we construct a Temporal Feature Abnormal Attention~(TFAA) module built from reconstruction learning and Cross-Reconstruction Attention Transformer~(CRATrans) block. The reconstruction learning can be used to determine abnormal states of multi-sensor time-series signals\cite{DBLP:journals/tkde/ZhangCWP23}. We believe that temporal features from different modalities can also be viewed as a type of multi-source data. We try to use an encoder-decoder structure to learn the distribution of real samples during the training phase. During the inference phase, we use an attention mechanism to focus on the abnormal segments generated by feature reconstruction, which can adapt well to the features extracted from various modality data. The proposed module is shown in Figure ~\ref{model_TFAA}.

\textbf{Reconstruction Learning.} To be specific, given the encoding feature sequence $F$, we first employ a Deep Convolutional AutoEncoder~(DCAE) as illustrated in Figure~\ref{model_TFAA}(a) to learn robust representations for real samples. The DCAE consists of a convolutional encoder $f_E$ and a de-convolutional decoder $f_D$. The encoder is composed of $L$ convolutional modules. Each convolution module contain a convolution layer followed by LeakyReLU, and Instance Normalization~\cite{DBLP:journals/corr/UlyanovVL16}. The low-dimensional representation $Z \in \mathbb{R} ^{C_{z} \times \frac{T}{2^L}}$ and the reconsturcted features $\hat{F}$ can be formulated as follows:

\begin{equation}
	\left\{
	\begin{matrix}
		Z = f_{E}\left(F\right), \\
		\hat{F} = f_{D}\left(Z\right).
	\end{matrix}
	\right.
\end{equation}
The $f_{E}$ encodes the input features into low-dimensional through convolution layers with a stride of 2. $Z$ represents the latent distribution of real samples. The $f_{D}$ decodes the latent low-dimensional representation to reconstruct the feature. The decoder is composed of transpose convolution layer, activation layer, and normalization layer.

During the training, we compute the distance between input features and reconstructed features of unmodified samples in a mini-batch as:

\begin{equation}
L_{rec} = {\frac{1}{N_r} \textstyle \sum_{i}^{N_r}} \Vert\hat{F}_i-F_i\Vert_1,
\end{equation}
where $N_r$ is the number of unmodified samples in a mini-batch, and $\Vert \cdot \Vert_1$ is the $l_1$-norm.
To enhance the consistency of real samples in the low-dimensional embedding space, we utilize a sample-level classifier, denoted as $f_S$, to distinguish the category to which the current feature sequence belongs - whether it is real or tampered. The classifier $f_S$ extracts sample-level features from latent features $Z$ using average pooling and passes them through two fully connected layers to obtain the probability score $p_t$ for the sample being tampered. To address the issue of imbalanced data between real and tampered samples, we utilize the focal loss~\cite{DBLP:conf/iccv/LinGGHD17} as the loss function during training. The sample-level focal loss is computed as follows:

\begin{equation}
L_{scls} = -\alpha(1-p_t)^\gamma \log(p_t),
\end{equation}
where $\alpha$ is weighting factor to balance positive and negative samples and $\gamma$ is the modulating factor to balance easy and hard samples. 

\textbf{Cross-Reconstruction Attention Transformer.} Furthermore, many existing anomaly detection algorithms for time series data use reconstruction error to identify abnormal segments. These algorithms set a threshold and flag any segments with reconstruction error above the threshold as anomalous. However, for our task, we need to consider the difference in information carried by different types of samples, which can affect the difficulty of reconstruction and lead to larger errors in some real samples. Additionally, manipulated segments can be very similar to real segments, resulting in small differences in reconstruction. Therefore, directly using reconstruction error to improve our algorithm's performance is difficult. 

To address above problem, we introduce a CRATrans module, as shown in Figure ~\ref{model_TFAA}(b) . As mentioned in ~\cite{DBLP:conf/eccv/ZhangWL22}, transformer block with self-attention module computes a weighted average of features by assigning weights proportional to the similarity score between pairs of input features. In our case, our CRATrans block with Cross-Reconstruction Attention (CRA) will compute similarity scores between pairs of original and reconstructed features in order to replace simple reconstruction errors. 

In detail, given the original features $F \in \mathbb{R}^{C \times T}$
and reconstructed features $\hat{F} \in \mathbb{R}^{C \times T}$, we add positional encodings~\cite{DBLP:conf/nips/VaswaniSPUJGKP17} at these features to make position-sensitive feature $F_{pe}$ and $\hat{F}_{pe}$. We believe that positional encodings help to enhance the attention to subtle changes in temporal features. Then we transform them into a latent space by using Layer Normalization~(LN)~\cite{DBLP:journals/corr/BaKH16} and learnable parameter matrices $\left\{W_q, W_k\right\} \in \mathbb{R}^{D \times C}$, respectively. The query $F_{q}$ and key $F_{k}$ are calculated by
\begin{equation}
F_{q} = W_{q}\left(LN\left(F_{pe}\right)\right), F_{k} = W_{k}\left(LN\left(\hat{F}_{pe}\right)\right),
\end{equation}
where $\left \{F_{q},F_{k}\right \} \in \mathbb{R}^{D \times T}$. The original-reconstructed correlation matrix $F_{r} \in \mathbb{R}^{T \times T}$is given by
\begin{equation}
F_{r} = F_{q}^\top F_{k},
\end{equation}
which represents the similarity between the original features and the reconstructed features in the temporal domain. A CRA matrix $F_{cra}$ is obtained by normalizing the correlation matrix  $F_{r}$, as follows:
\begin{equation}
F_{cra} = {Softmax}\left(\frac{F_{r}}{\sqrt{C}} \right),
\end{equation}
where ${Softmax}$ is performed row-wise, $\frac{1}{\sqrt{C}}$ is used as the scaling factor. This approach can effectively avoid misjudgment or neglect of abnormalities between the reconstructed and original features due to factors such as scale. Meanwhile, we project the feature $F_{pe}$ to value $F_{v} \in \mathbb{R}^{D \times T}$ by using the LN and learnable parameter matrix $W_{v}$:
\begin{equation}
F_{v} = W_{v}\left(LN\left(F_{pe}\right)\right).
\end{equation}

In next step, a dot-product is performed on $F_{cra}$ and the feature $F_{v}$ to get the representation $F_{ea}$ enhanced by reconstruction anomaly attention. We formulate the function as
\begin{equation}
F_{ea} = F_{cra}F_{v},
\end{equation}
where $F_{ea} \in \mathbb{R}^{D \times T}$. Furthermore, we actually used a Multi-head Cross-Reconstruction Attention(MCRA) for our model, where several CRA operations are concatenated together in parallel.

The output features $F_{ea}$ are added to the original feature $F_{pe}$ and are normalized by the LN layer. Finally, We employ a simple fully connected feed-forward network~(FFN) with a residual connection to product the output $F_{crat}$ of CRATrans block.


\subsection{Parallel Cross-Attention Feature Pyramid Network}
\begin{figure}[htbp]
    \centering
    \subfigure[Hierarchical Network]{\label{fig:fpn:a}\includegraphics[width=0.3\columnwidth]{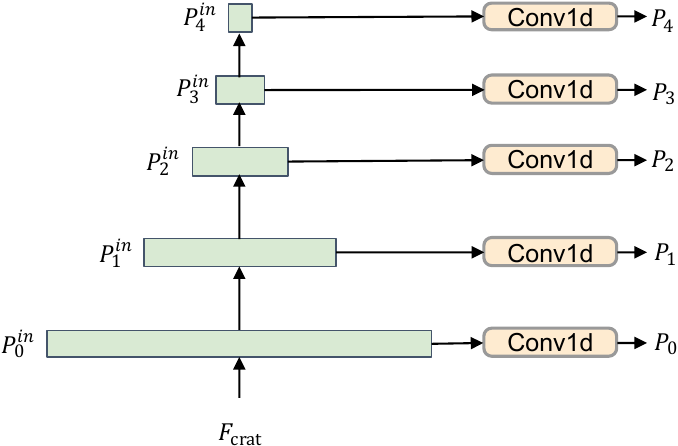}} 
    \subfigure[FPN]{\label{fig:fpn:b}\includegraphics[width=0.5\columnwidth]{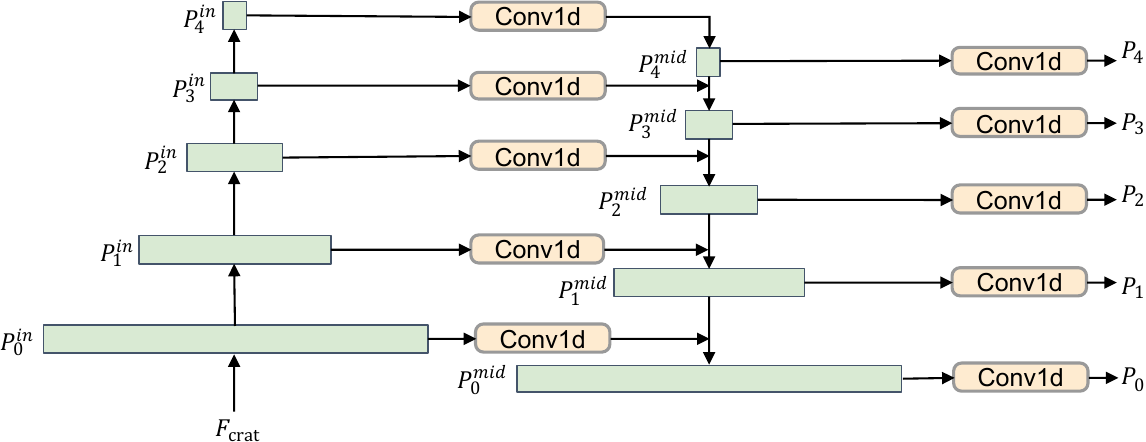}} \\
    \subfigure[PCA-FPN]{\label{fig:fpn:c}\includegraphics[width=0.8\columnwidth]{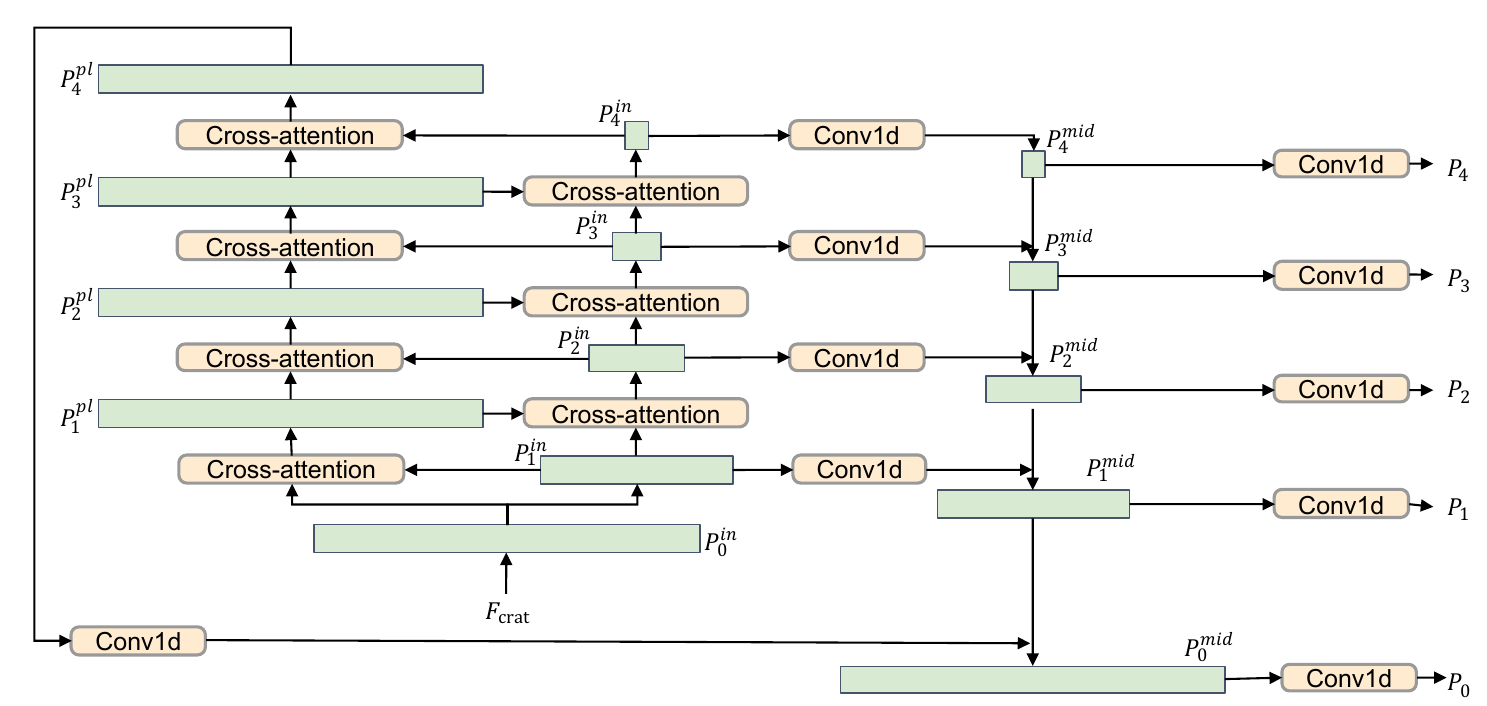}}
    \caption{Comparison of feature pyramid networks design in the case of 5 levels.}
    \label{fig:fpn}
\end{figure}

High-resolution feature maps are crucial for position-sensitive tasks, such as TFL, which involve numerous short video segments. A multi-scale Transformer encoder was used in ~\cite{DBLP:conf/eccv/ZhangWL22} to locate action segments in video based on features maps of different resolutions. This encoder utilizes a simple hierarchical multi-scale network, as shown in Figure~\ref{fig:fpn:a}. However, the limited representation capability of high-resolution feature maps for complex content poses a challenge. To address this issue, ~\cite{DBLP:conf/cvpr/LinDGHHB17} is commonly used to fuse features of different scales to improve the network's temporal localization ability. The scheme of FPN is shown in Figure~\ref{fig:fpn:b}. Despite its effectiveness, the fusion process using a simple form of upsampling and downsampling followed by addition usually introduces noise to features of different levels, which may interfere with localization. This effect is particularly pronounced for shorter segments, where even small localization deviations can cause a sharp change in the temporal Intersection over Union (tIoU) between predicted and true values. For example, a segment of 0.5 seconds, when shifted by 0.1 seconds from its correct position, can result in a $20\%$ decrease in tIoU, while for 2 seconds, the tIoU will only decrease by $5\%$. Inspired by HRNet~\cite{DBLP:journals/pami/00010CJDZ0MTW0X21}, we propose a Parallel Cross-Attention Feature Pyramid Network~(PCA-FPN) to enhance high-resolution features in such cases. The PCA-FPN is illustrated in Figure~\ref{fig:fpn:c}, and effectively addresses the problem of noise in feature fusion, improving the localization performance of the network.

The PCA-FPN fuses features of different scales simultaneously through parallel and down-sampling branches, and improves their interaction through a cross-attention~(CA) mechanism, Specifically, given the input feature $F_{crat}$ from TFAA module we can encode it to obtain a high-resolution feature map, denoted as $P_{0}^{in} \in \mathbb{R}^{D_{fpn} \times T}$. Similar to other methods, $P_{0}^{in}$ is downsampled by an encode module with a factor of 2 to obtain a medium-resolution feature map $P_{1}^{in} \in \mathbb{R}^{D_{fpn} \times \frac{T}{2}}$. Following ~\cite{DBLP:conf/eccv/ZhangWL22}, the encoder module is a multi-scale transformer unit. To further enhance the representation of the high-resolution feature map $P_{0}^{in}$, we feed these two different resolution feature maps $P_{0}^{in}$ and $P_{1}^{in}$ into the CA module to enhance their features. The CA module is calculated as follows:
\begin{equation}
\small
P_{1}^{pl} = CA\left(W_{cq}\left(LN\left(P_{0}^{in}\right)\right),W_{ck}\left(LN\left(P_{1}^{in}\right)\right),W_{cv}\left(LN\left(P_{1}^{in}\right)\right)\right),
\end{equation}
where 
\begin{equation}
CA \left(Q,K,V\right) = {Softmax}\left(\frac{Q^\top g\left(K\right)}{\sqrt{D_{fpn}}} \right)g\left(V\right),
\end{equation}
$\left\{W_{cq}, W_{ck}, W_{cv}\right\}\in \mathbb{R}^{D_{fpn} \times D_{fpn}}$ are learnable parameter matrices, $g\left ( \cdot \right )$ is temporal interpolation function that resamples the $K$ and $V$, which are the inputs of CA to the same size as $Q$, ${Softmax}$ is performed row-wise, $\frac{1}{\sqrt{D_{fpn}}}$ is used as the scaling factor and $P_{1}^{pl}$ is the first level parallel high-resolution feature. Subsequently, Subsequently, $P_{1}^{in}$ used as Query $Q$ and $P_{1}^{pl}$ used as Key $K$ and Value $V$ to CA module. The output of the CA module is then passed to the multi-scale transformer unit for downsampling to obtain $P_{2}^{in}$. These processes preserve the feature of short segments in the high-resolution feature map while enhancing the representation of features at different scales. By repeating these processes, we can obtain five levels of parallel multi-scale features $\left\{ P_{4}^{pl}, P_{1}^{in}, P_{2}^{in}, P_{3}^{in}, P_{4}^{in}\right\}$. Finally, we fuse the five levels of features from top to bottom, similar to FPN, to obtain the enhanced multi-scale features $P = \left\{ P_{0}, P_{1}, P_{2}, P_{3}, P_{4}\right\}$.

\subsection{Training and Inference}
The given feature pyramid $P$ can be decoded into output $S=\left\{t_{n,sf},t_{n,ef}, s_{n,f}\right\}_{n=1}^{N_f}$ through classification and regression heads. The final training loss for the overall model is:
\begin{equation}
L = L_{cls} + \lambda_{reg} L_{reg} + \lambda_{rec} L_{rec} + \lambda_{scls} L_{scls},
\end{equation}
where $L_{cls}$ and $L_{reg}$ are losses for the classification head outputs $\left\{s_{n,f}\right\}_{n=1}^{N_f}$ and regression head outputs $\left\{t_{n,sf},t_{n,ef}\right\}_{n=1}^{N_f}$, respectively. $L_{cls}$ is binary classification loss, where the label of forgery segments is set to 1 and the rest is set to 0. Other settings are directly adopted from ActionFormer. The reconstruction loss  $L_{rec}$ and sample-level focal loss $L_{scls}$ are mentioned in section~\ref{sec:TFAA}. $\lambda_{reg}$, $\lambda_{rec}$ and $\lambda_{scls}$ are hyper-parameters used to balance the relationship between the losses. By default, we set $\lambda_{reg}=2$, $\lambda_{rec}=1$, and $\lambda_{scls}=0.1$.

For the inference stage, we applied Soft-NMS~\cite{DBLP:conf/iccv/BodlaSCD17} to post-process the results and remove a large number of redundant predictions.

\section{Temporal Video Inpainting Localization}
With the rapid development of AIGC technology, highly deceptive video and audio content has been widely spread on the Internet, leading to potential harm caused by the spread of misleading information. While benchmarks for deepfake videos~\cite{DBLP:conf/nips/KhalidTKW21,DBLP:journals/corr/abs-1910-08854,DBLP:conf/mm/ZiCCMJ20,DBLP:conf/iccv/RosslerCVRTN19} and audios~\cite{DBLP:conf/nips/FrankS21} have emerged in recent years to address the forgery of facial or speech content, these methods only cover a small portion of all forged content. There is a lack of relevant dataset research for other harmful forgery methods. Therefore, we synthesized a dataset for locating video inpainting segments as a new benchmark for TFL, namely TVIL. Our goal is to detect various types of inpainting forgery in sequential images or videos to defend against the spread of misinformation and bring new insights to the research community.

\textbf{Data Collection.} The dataset is constructed based on YouTube-VOS 2018~\cite{DBLP:conf/eccv/XuYFYYLPCH18}, which contains over 4,000 online videos from YouTube. Considering that YouTube is currently one of the most popular video platforms and also an important source for generating and spreading misleading information, we believe that generating a synthesized dataset based on YouTube videos can effectively evaluate the performance of TFL algorithms and prevent the spread of misinformation.

\textbf{Data Processing.} YouTube-VOS 2018 is a semi-supervised video semantic segmentation dataset that does not provide complete segmentation masks required for video inpainting. Therefore, we utilized XMEM~\cite{DBLP:conf/eccv/ChengS22}, a state-of-the-art video semantic segmentation algorithm, to generate the segmentation masks. These generated masks can be classified into two types: stationary masks and moving masks~\cite{DBLP:conf/eccv/ZengFC20}, which are widely used in real-world scenarios. Stationary masks can be used for removing static objects, simulating the removal of visible watermarks leading to copyright infringement, and so on. On the other hand, moving masks can be used for removing moving objects, simulating the removal of specific targets such as people in surveillance videos. This technology can potentially be used to provide false evidence in certain situations. To better simulate real-world scenarios, we randomly split the dataset into five parts, where one part is used as the real sample set without any manipulation. The remaining four parts are subjected to different video inpainting methods, namely STTN~\cite{DBLP:conf/eccv/ZengFC20}, FuseFormer~\cite{DBLP:conf/iccv/0019DHSLS0D021}, E2FGVI~\cite{DBLP:conf/cvpr/0031LQGC22} and FGT~\cite{DBLP:conf/eccv/ZhangFL22}, which randomly removed some frames of the target object. This process aimed to create more diverse and challenging samples to test the effectiveness of the proposed method in handling complex scenarios.

\begin{figure}[htbp]
    \centering
    \subfigure[train set]{\label{fig:video_len:a}\includegraphics[width=0.32\columnwidth]{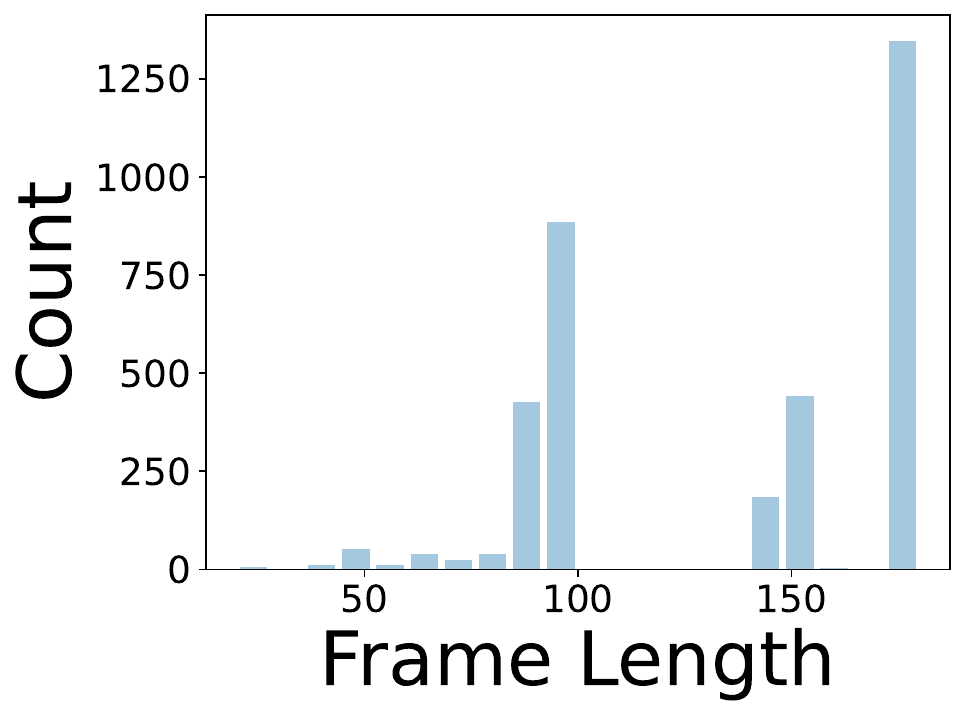}}
    \subfigure[validation set]{\label{fig:video_len:b}\includegraphics[width=0.32\columnwidth]{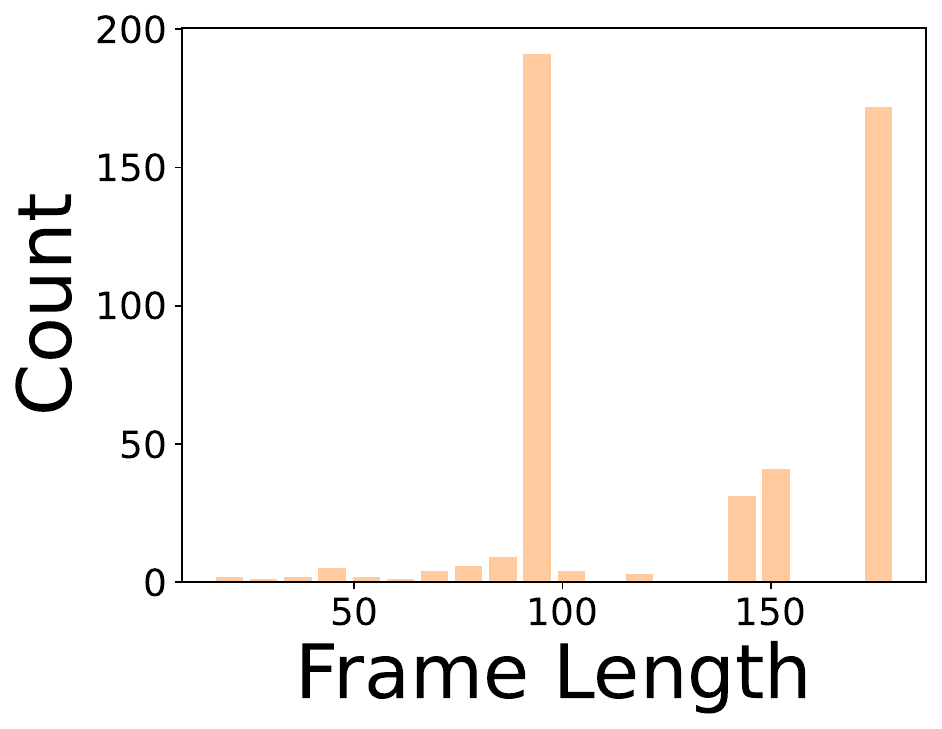}}
    \subfigure[test set]{\label{fig:video_len:c}\includegraphics[width=0.32\columnwidth]{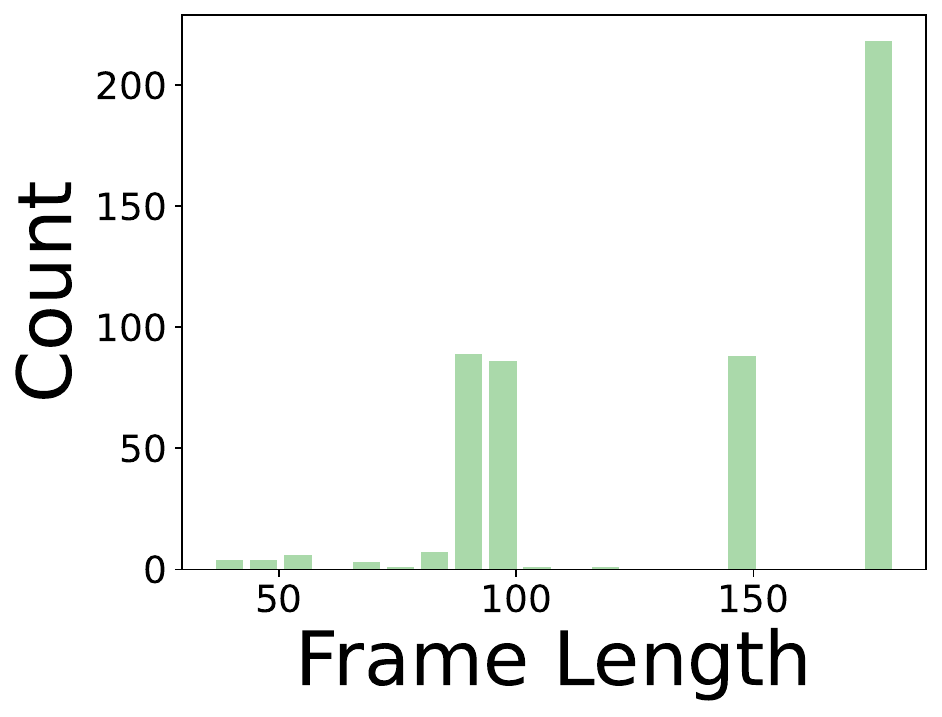}}
    \caption{Distribution of video lengths.}
    \label{fig:video_len}
\end{figure}

\textbf{Dataset Statistics.} We follow the original split in YouTube-VOS 2018, which consisted of 3,471 video clips for training, 474 for validation and 508 for testing. The average length of video clips is about 140 frames, as shown in Figure~\ref{fig:video_len}. The training set consists of 3340 forgery segments, the validation set consists of 451 forgery segments and the test set consisists of 463 forgery segments. In our task, video clips with a duration of less than 1 second are defined as short clips. Compared to the Lav-DF~\cite{DBLP:journals/corr/abs-2204-06228} dataset, where $89.26\%$ of the manipulated clips are short, our dataset has a proportion of $99.60\%$, making our dataset more challenging. The distribution of our dataset is illustrated in Figure~\ref{fig:video_dist}. In Appendix~\ref{app:a1}, we provide a further comparison between the TVIL dataset and other commonly used multimedia forensic datasets.

\begin{figure}[htbp]
    \centering
    \subfigure[Inpainting Methods]{\label{fig:video_dist:a}\includegraphics[width=0.30\columnwidth]{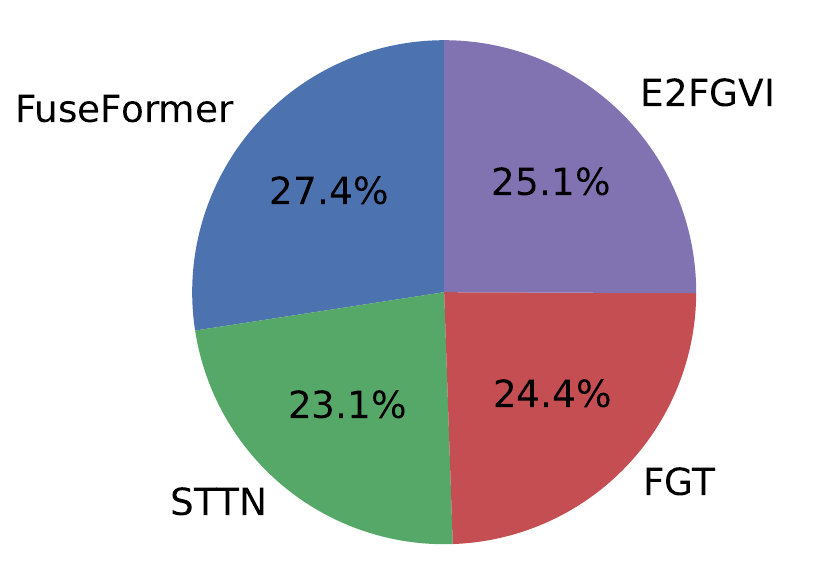}}
    \subfigure[No. of Fake Segments]{\label{fig:video_dist:b}\includegraphics[width=0.30\columnwidth]{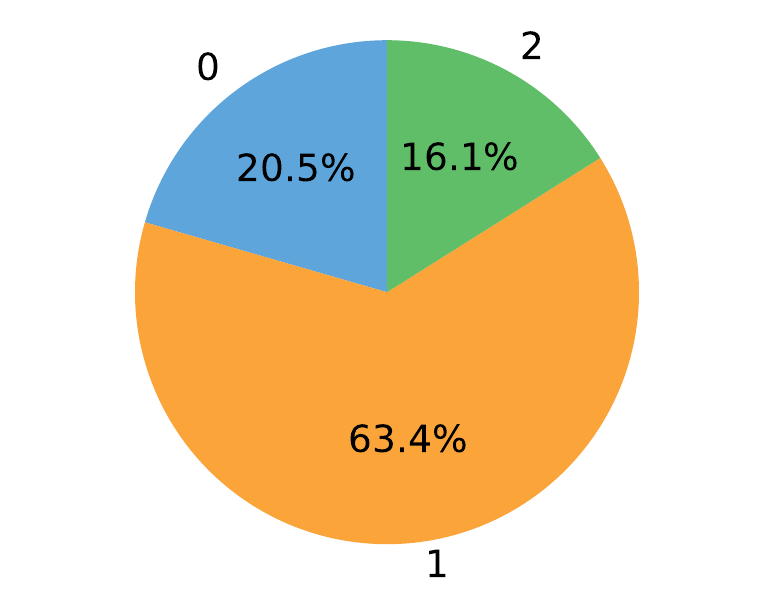}}
    \subfigure[Fake Segments Ratio]{\label{fig:video_dist:c}\includegraphics[width=0.30\columnwidth]{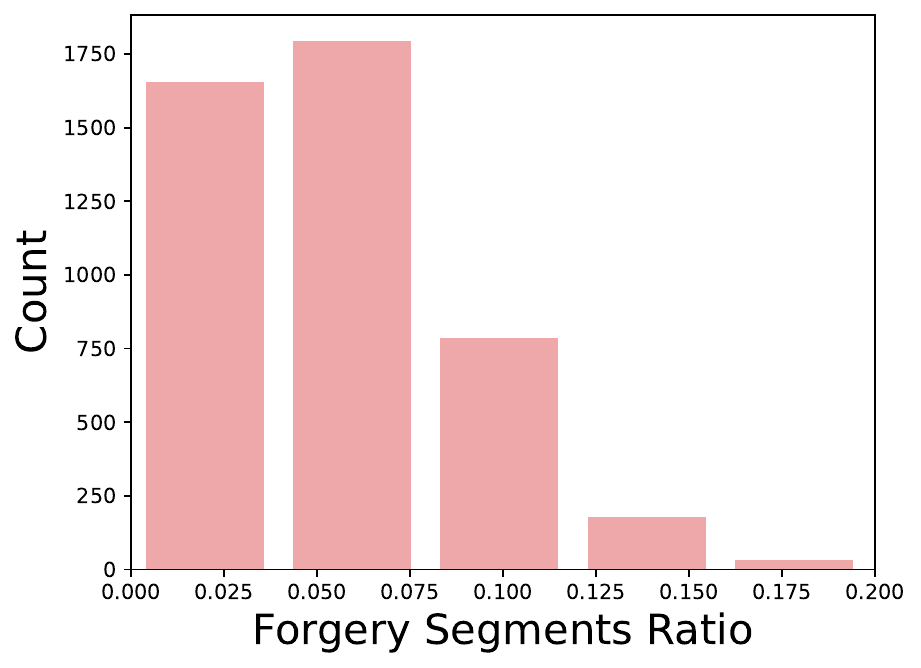}}
    \caption{Distribution of the TVIL datasets. (a) The ratio of different methods used in modified video segments. (b) The proportion of manipulated segments in the dataset. (c) The proportion of different manipulated clip lengths to the total length of the video.}
    \label{fig:video_dist}
\end{figure}

\section{EXPERIMENTS}

\subsection{Experimental Setup.} For visual data, we use the two-stream TSN~\cite{DBLP:conf/eccv/WangXW0LTG16} network pre-trained on ActivityNet dataset~\cite{DBLP:conf/cvpr/HeilbronEGN15} to extract the two-stream visual features. The optical flow is extracted by TV-$L1$ algorithm. The frame interval is set to 1. For audio data, we employ a pre-trained BYOL-A~\cite{DBLP:conf/ijcnn/NiizumiTOHK21} pre-trained on AudioSet~\cite{DBLP:conf/icassp/GemmekeEFJLMPR17}. The dimension of the extracted video features is 4096, while that of the audio features is 2048. The extracted features are interpolated to 768 in the temporal dimension. 

\textbf{Datasets and Evaluation Metric.} We evaluate our method on three benchmark datasets, including Lav-DF~\cite{DBLP:journals/corr/abs-2204-06228} for multi-modal data in face forgery scenarios, our proposed TVIL dataset for visual modality in general scenes beyond faces, Psynd~\cite{DBLP:conf/icpr/ZhangS22} for audio modality data in speech scenarios. We follow the evaluation protocol in \cite{DBLP:conf/cvpr/HeGCZYSSS021,DBLP:journals/corr/abs-2204-06228} and report average precision (AP) and average recall (AR) as evaluation metrics, Following conventions, we set the tIoU threshold values at $\left\{0.5,0.75, 0.95 \right\}$ and set Average Number of proposals~(AN) to $\left\{10,20,50,100\right\}$. In addition, for dataset Psynd, we also provide tIoU to follow the protocol of dataset Psynd.

\textbf{Baseline and Comparison.}
We use ActionFormer~\cite{DBLP:conf/eccv/ZhangWL22} as our baseline network and reproduced it based on the official
code\footnote{\url{https://github.com/happyharrycn/actionformer_release}} with default settings on our own datasets. We extend the advanced TAL network, DCAN~\cite{DBLP:conf/aaai/ChenZWL22} and TAGS~\cite{DBLP:conf/eccv/NagZSX22}, on the TVIL dataset, representing research efforts focused on enhancement of boundary features and improvement of head location, respectively. Additionally, we compare our algorithm with the state-of-the-art methods on each dataset to quantitatively evaluate the performance of our approach.

\textbf{Implementation Details.} We follow ActionFormer with minor modifications as follows. Our models are trained on a single RTX 3090 GPU with initial learning rate of 0.001. The batch size for Lav-DF is 32, for TVIL is 16 and for Psynd is 8. 

\subsection{Results for Temporal Face Forgery Localization}
We report the AP and AR performance of our method and state-of-the-art methods on the Lav-DF Full Set in Table~\ref{tab:lavdf_full}. For the full set, which includes three types of attacks (audio-only modified, video-only modified, and audio-video modified), most unimodal models~\cite{DBLP:conf/mm/ChughGDS20, DBLP:journals/corr/abs-2101-08540, DBLP:conf/iccv/LinLLDW19} that focus only on visual information struggle to accurately locate the tampered segments. Although multimodal models~\cite{DBLP:conf/visapp/BagchiMFS22, DBLP:journals/corr/abs-2204-06228} perform well in terms of AP at tIoU 0.5, they completely fail for the more challenging AP at tIoU 0.95. The main reason for this is the lack of effective feature enhancement for short video segments. Short segments are extremely sensitive to the tIoU metric. ActionFormer~\cite{DBLP:conf/eccv/ZhangWL22} network introduces a simple hierarchical transformer-based network that effectively improves both AP and AR. Furthermore, our method further outperforms BA-TFD\cite{DBLP:journals/corr/abs-2204-06228} by $37.36\%$ in terms of AP at tIoU 0.95 through the proposed PCA-FPN and TFAA with mutilmodal features. For visual-only feature as inputs, our significantly improves AP at tIoU 0.95 from $0.16\%$ to $25.68\%$ compare with BA-TFD. In Appendix~\ref{appendix:lavdf-subset}, we further present the experimental results for the Lav-DF subset. In Appendix~\ref{appendix:lavdf-classification}, we provide the forgery classification results and additional evaluation metrics for the Lav-DF Full Set.

\begin{table}[htbp]
\centering
\caption{Performance comparison on Lav-DF Full Set. Bold faces correspond to the top performance.}
\label{tab:lavdf_full}
\resizebox{\columnwidth}{!}{%
\begin{tabular}{cc|ccccccc}
\hline
\multirow{2}{*}{Methods} & \multirow{2}{*}{Feature} & \multicolumn{7}{c}{Full Set}                                               \\ \cline{3-9} 
                         &                          & AP@0.5    & AP@0.75   & AP@0.95   & AR@10     & AR@20     & AR@50     & AR@100 \\ \hline
MDS~\cite{DBLP:conf/mm/ChughGDS20}   & Visual       & 12.78     & 1.62      & 0.00      & 37.88     & 36.71     & 34.39     & 32.15  \\
AGT~\cite{DBLP:journals/corr/abs-2101-08540} & Visual                   & 17.85     & 9.42      & 0.11      & 43.15     & 34.23     & 24.59     & 16.71  \\
BMN~\cite{DBLP:conf/iccv/LinLLDW19}                      & Visual                   & 24.01     & 7.61      & 0.07      & 53.26     & 41.24     & 31.60      & 26.93    \\
BMN (I3D)~\cite{DBLP:conf/iccv/LinLLDW19}                & Visual                   & 10.56     & 1.66      & 0.00      & 48.49     & 44.39     & 37.13     & 31.55    \\
AVFusion~\cite{DBLP:conf/visapp/BagchiMFS22}                 & Visual+Audio             & 65.38     & 23.89     & 0.11      & 62.98     & 59.26     & 54.80     & 52.11     \\
\multirow{2}{*}{BA-TFD~\cite{DBLP:journals/corr/abs-2204-06228}}  & Visual                   & 58.55     & 28.60     & 0.16      & 62.49     & 58.77     & 53.86     & 50.29    \\
                         & Visual+Audio             & 76.90      & 38.50     & 0.25      & 66.90     & 64.08     & 60.77     & 58.42        \\
ActionFormer~\cite{DBLP:conf/eccv/ZhangWL22}             & Visual                   & 95.34     & 90.20     & 23.73     & 88.41     & 89.63     & 90.33     & 90.41      \\ \hline
\multirow{2}{*}{Ours}    & Visual       &97.30      &92.96      &25.68      &90.19      &90.85      &91.14       &91.18  \\
                         & Visual+Audio  &\textbf{98.83} &\textbf{95.54} &\textbf{37.61} &\textbf{92.10} &\textbf{92.42} &\textbf{92.47} &\textbf{92.48}	\\ \hline
\end{tabular}%
}
\end{table}

\subsection{Results for Temporal Video Inpainting Localization}
Experimental results in Table~\ref{tab:tvil_full} show that our method outperforms all compared TAL methods on both the AP and AR evaluations on the proposed TVIL dataset. DCAN~\cite{DBLP:conf/aaai/ChenZWL22} is a boundary-enhanced algorithm based on BMN~\cite{DBLP:conf/iccv/LinLLDW19} implementation. TAGS~\cite{DBLP:conf/eccv/NagZSX22} is a based on a novel localization head that does not include a regression task. It is worth mentioning that due to the increase in in the number of short video clips, the overall performance of ActionFormer is lower than that of Lav-DF. Nevertheless, our method still achieved the best performance, showing the superiority of our method.

\begin{table}[htbp]
\caption{Comparison between our method and other state-of-the-art TAL methods on TVIL. Bold faces correspond to the top performance.}
\label{tab:tvil_full}
\resizebox{\columnwidth}{!}{%
\begin{tabular}{cccccccc}
\hline
Methods      & AP@0.5               & AP@0.75              & AP@0.95              & AR@10                & AR@20                & AR@50                & AR@100               \\ \hline
TAGS~\cite{DBLP:conf/eccv/NagZSX22} & 18.40       & 12.68    & 0.09               & 24.41               & 25.05                & 25.56                & 25.56              \\
DCAN~\cite{DBLP:conf/aaai/ChenZWL22} & 82.75       & 75.00    & 3.22               & 64.73               & 66.02                & 68.82                & 69.97              \\
ActionFormer~\cite{DBLP:conf/eccv/ZhangWL22} & 86.27       & 83.03                & 28.17                & 84.82                & 85.77                & 88.10                & 88.49                \\ \hline
Ours         &\textbf{88.68} &\textbf{84.70} &\textbf{62.43} &\textbf{87.09} &\textbf{88.21} &\textbf{90.43} &\textbf{91.16} 	   \\ \hline
\end{tabular}%
}
\end{table}

\subsection{Results for Partial Synthetic Speech Localization}
We further evaluate UMMAFormer on partial synthetic speech localization task to illustrate its superiority on different modal adaption. Following LFSS~\cite{DBLP:conf/icpr/ZhangS22}, we report tIoU on Psynd dataset. LFSS is the only work so far focused on localizing voice cloning partially faked English
speech. Due to the absence of completely real samples in the original training set of Psynd, we randomly extracted 299 unaltered audio segments from the original dataset as the real training samples for our proposed method. To ensure fairness, we also selected the best tIoU at different thresholds as the final result, as LFSS did. As shown in Table~\ref{tab:Psynd_full}, our algorithm achieves better performance under most conditions, especially in the landline and cellular test subsets. This means that our algorithm has good robustness even when the audio is further disturbed. It is worth mentioning that the test special subset contains both completely fake and completely real samples. Under this condition, LFSS based on simple binary classification achieves good results and outperforms our method by $1.11\%$. However, for other more challenging scenarios, such as local partial modified segments, especially under conditions where data are disturbed, our proposed method performs better. This further demonstrates the value of research on TFL tasks. 

\begin{table}[htbp]
\caption{Performance comparison on Psynd in terms of tIoU. Bold faces correspond to the top performance.}
\label{tab:Psynd_full}
\resizebox{0.7\columnwidth}{!}{%
\begin{tabular}{ccccc}
\hline
Methods       & test set & special test set & landline & cellular \\ \hline
LFSS~\cite{DBLP:conf/icpr/ZhangS22}         & 98.58    & \textbf{99.35}            & 80.29    & 80.94  \\ \hline
Ours         & \textbf{98.70}    & 98.24            & \textbf{92.04}   & \textbf{86.57}     \\ \hline
\end{tabular}%
}
\end{table}

\subsection{Ablation Studies}

We compare the contributions of different components of our method for different modalities. Table~\ref{tab:Ablation_module_FPN} shows the comparison between the performance of our proposed PCA-FPN and FPN. The experiments are conducted under three scenarios: visual-audio~(Lav-DF Full Set), visual-only~(TVIL), and audio-only~(Psynd-Test). The baseline refers to ActionFormer. We observed that FPN actually reduces model performance in audio tasks because it introduces noise during the multi-scale fusion process. On the other hand, our proposed PCA-FPN greatly improves the localization accuracy of the model. We also find that PCA-FPN can be applied to localization tasks in different modalities.

\begin{table}[htbp]
\centering
\caption{Comprasion with FPN. Bold faces correspond to the top performance of each dataset.}
\label{tab:Ablation_module_FPN}
\resizebox{0.9\columnwidth}{!}{%
\begin{tabular}{ccccccccc}
\hline
Dataset                  & Methods           & AP@0.5 & AP@0.75 & AP@0.95 & AR@10 & AR@20 & AR@50 & AR@100 \\  \hline
\multirow{3}{*}{Lav-DF Full Set}    
                        &Baseline        &97.58 &93.75 &40.38 &92.23 &92.71 &92.87 &92.90 	   \\
                        &Baseline+FPN   &\textbf{98.84} &\textbf{95.61} &38.63 &92.30 &92.59 &\textbf{92.65} &\textbf{92.66} 	   \\
                        &Baseline+PCA-FPN   &98.72 &95.52 &\textbf{39.00} &\textbf{92.31} &\textbf{92.60} &\textbf{92.65} &\textbf{92.66}         \\ \hline
\multirow{3}{*}{TVIL}   &Baseline   &86.10 &82.86 &28.11 &84.68 &85.71 &88.04 &88.43 	 	   \\ 
                        &Baseline+FPN     &88.50 &84.35 &38.95 &\textbf{85.91} &87.26 &\textbf{89.63} &\textbf{90.09}   \\
                        &Baseline+PCA-FPN   &\textbf{88.57} &\textbf{84.82} &\textbf{40.37} &85.56 &\textbf{87.44} & 89.53 &89.78	         \\ \hline 
\multirow{3}{*}{Psynd-Test}  &Baseline   &\textbf{100.00} &\textbf{100.00} &71.08 &95.95 &95.95 &95.95 &95.95 \\
                              &Baseline+FPN       &43.28 &5.13 &0.11 &47.22 &48.48 &48.86 &48.86  \\ 
                              &Baseline+PCA-FPN   &\textbf{100.00} &98.54 &\textbf{77.72} &\textbf{97.34} &\textbf{97.34} &\textbf{97.34} &\textbf{97.34}        \\ \hline
\end{tabular}%
}
\end{table}

Table~\ref{tab:Ablation_module_TFAA} provides further evidence of the value of TFAA, which was evaluated in the same three scenarios as mentioned earlier. The results show that TFAA effectively improved the model's performance in most scenarios, suggesting that it enhances the applicability of different modality features. Additionally, our model demonstrates the capability of universal modality-adaptation.

\begin{table}[htbp]
\centering
\caption{Ablation studies of the proposed TFAA modules. Bold faces correspond to the top performance of each dataset.}
\label{tab:Ablation_module_TFAA}
\resizebox{0.9\columnwidth}{!}{%
\begin{tabular}{ccccccccc}
\hline
Dataset                          & Methods                           & AP@0.5          & AP@0.75         & AP@0.95        & AR@10          & AR@20          & AR@50          & AR@100         \\ \hline
\multirow{3}{*}{Lav-DF Full Set} & Baseline                          & 97.58           & 93.75           & 40.38          & 92.23          & 92.71          & 92.87          & 92.90          \\
                                 & Baseline+TFAA                     & 97.57           & 93.74           & \textbf{40.53} & \textbf{92.31} & \textbf{92.80} & \textbf{92.98} & \textbf{92.99} \\
                                 & Baseline+PCA-FPN+TFAA~(ours) & \textbf{98.83}  & \textbf{95.54}  & 37.61          & 92.10          & 92.42          & 92.47          & 92.48          \\ \hline
\multirow{3}{*}{TVIL}            & Baseline                          &86.10          &82.86             &28.11        &84.68          &85.71         &88.04            &88.43          \\
                                 & Baseline+TFAA                      &85.82          &83.23            &51.71         &86.32          &87.48         &89.31           & 89.55 	          \\
                                 & Baseline+PCA-FPN+TFAA~(ours)      &\textbf{88.68} &\textbf{84.70} &\textbf{62.43} &\textbf{87.09} &\textbf{88.21} &\textbf{90.43} &\textbf{91.16} 	 \\ \hline
\multirow{3}{*}{Psynd-Test}      & Baseline                          & \textbf{100.00} & \textbf{100.00} & 71.08          & 95.95          & 95.95          & 95.95          & 95.95          \\
                                 & Baseline+TFAA                     & \textbf{100.00} & 98.41           & 76.23          & 97.09          & 97.09          & 97.09          & 97.09          \\
                                 & Baseline+PCA-FPN+TFAA~(ours) & \textbf{100.00} & \textbf{100.00} & \textbf{79.87} & \textbf{97.60} & \textbf{97.60} & \textbf{97.60} & \textbf{97.60} \\ \hline
\end{tabular}%
}
\end{table}

\begin{figure}[htbp]
    \centering
    \subfigure[test set]{\label{fig:tiou_comparison:a}\includegraphics[width=0.45\columnwidth]{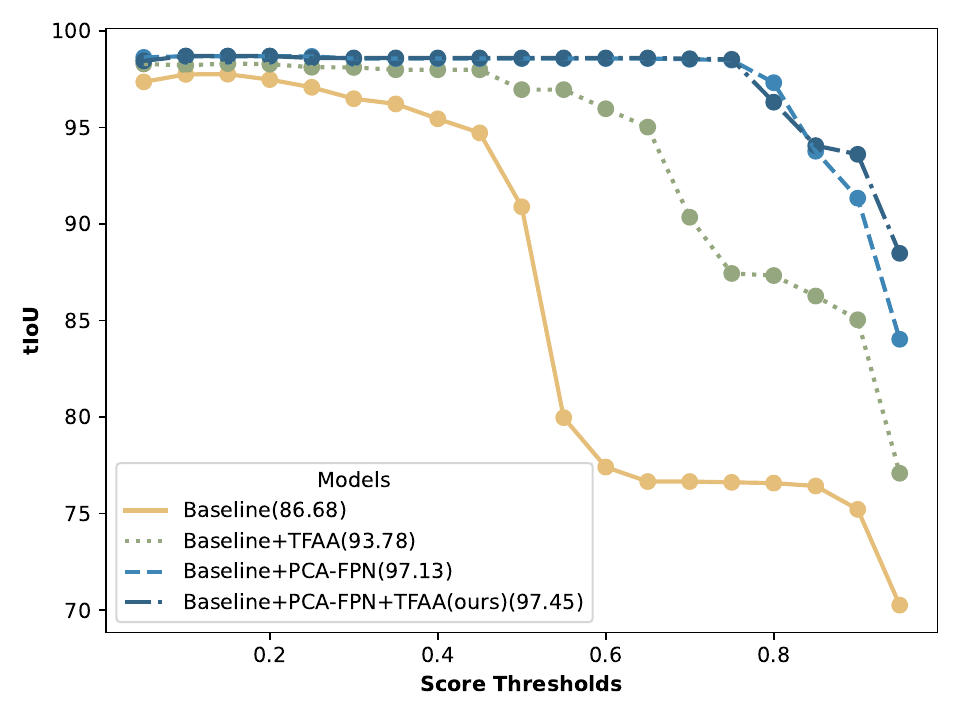}}
    \subfigure[special test set]{\label{fig:tiou_comparison:b}\includegraphics[width=0.45\columnwidth]{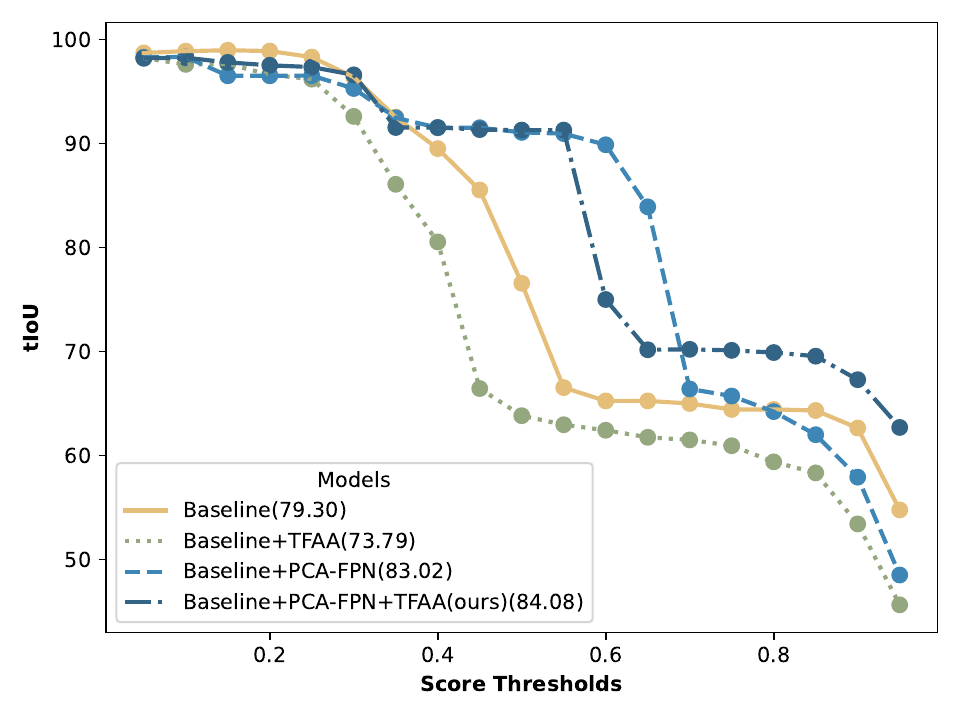}}\\
    \subfigure[landline]{\label{fig:tiou_comparison:c}\includegraphics[width=0.45\columnwidth]{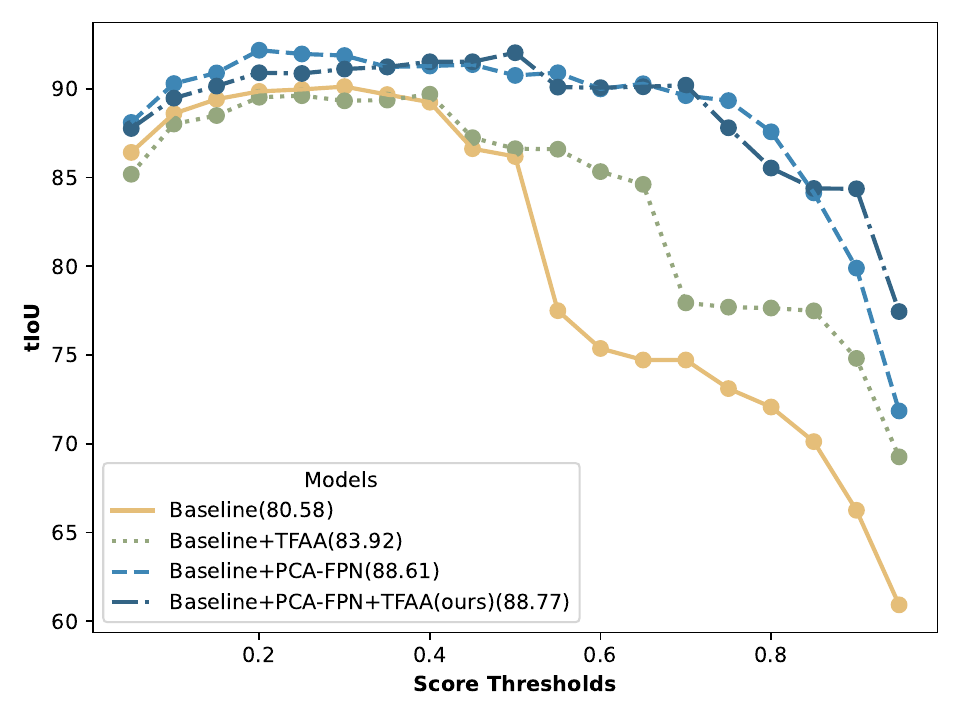}}
    \subfigure[cellular]{\label{fig:tiou_comparison:d}\includegraphics[width=0.45\columnwidth]{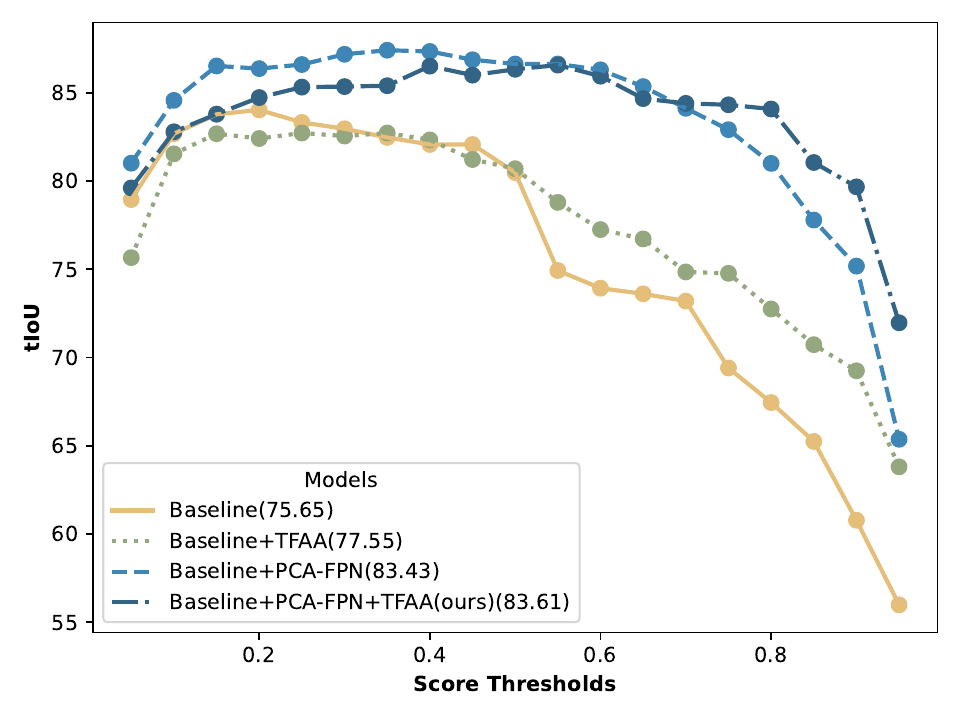}}
    \caption{Ablation studies with respect to effect of score thresholds on tIoU. Score threshold values varies from 0.05 to 0.95 with a step size of 0.05. We calculate the average tIoU over different thresholds.}
    \label{fig:tiou_comparison}
\end{figure}

Figure~\ref{fig:tiou_comparison} demonstrates the impact of different modules on tIoU. We conducted tests on Psynd, varying the confidence scores from 0.05 to 0.95, and compared the resulting tIoU values as well as their averages. A higher average value indicates a higher effectiveness of our predicted candidates. Both TFAA and PCA-FPN effectively improve the model's performance, and combining them results in even better performance.

\begin{figure}[btph]
    \centering
    \subfigure[Lav-DF]{\label{fig:samples:a}\includegraphics[width=1\columnwidth]{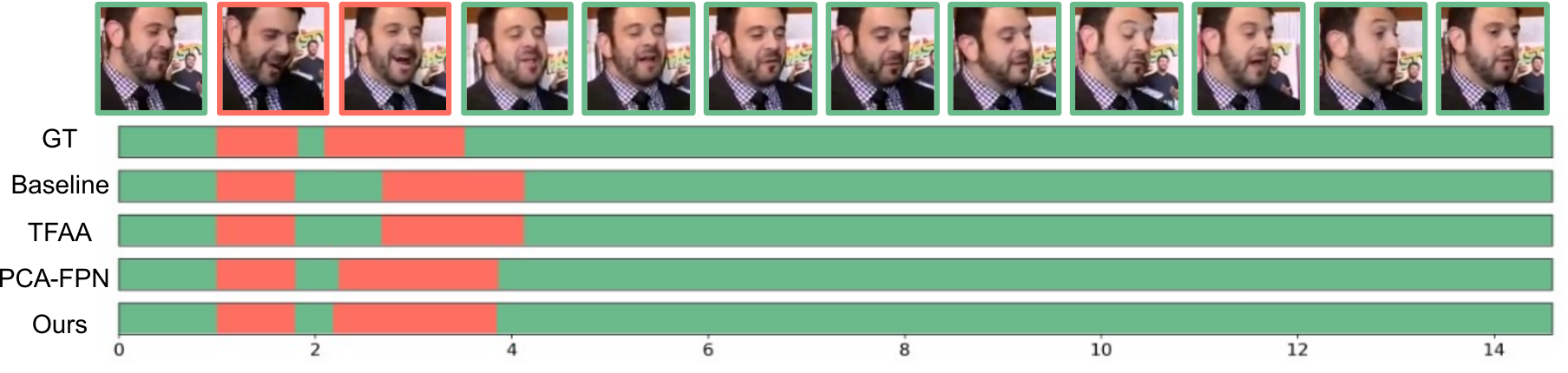}} \\
    \subfigure[TVIL]{\label{fig:samples:b}\includegraphics[width=1\columnwidth]{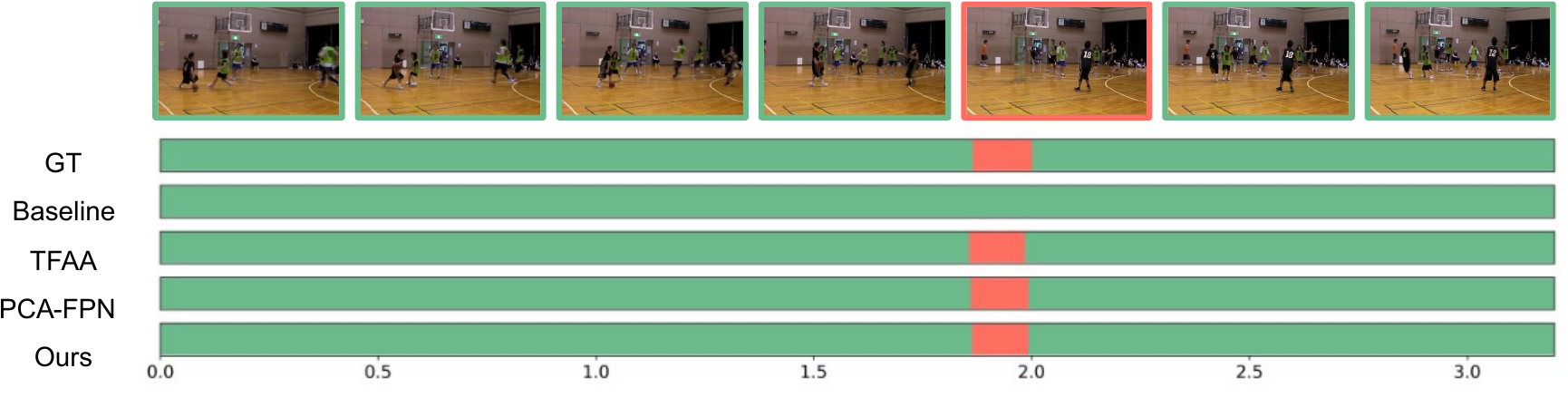}} \\
    \caption{Qualitative examples of our proposed model ablation experiments. Red indicates fake segments and green indicates real segments.}
    \label{fig:samples}
\end{figure}

Figure~\ref{fig:samples} provides visual representations of two qualitative examples from Lav-DF and TVIL. As shown in Figure~\ref{fig:samples:a}, the baseline method can locate the corresponding forged segments, but it exhibits a larger offset compared to our method. As depicted in Figure~\ref{fig:samples:b}, due to the difficulty in locating forged segments that belong to short segments, the baseline method failed to identify them, whereas our method achieved significant detection results.

\section{CONCLUSIONS}

In this paper, we propose a noval universal multimodal-adaptive transformer framework for TFL, which fosters deeper investigations in multimedia content security and helps prevent the misuse of AIGC. To solve the challenges in the task, we propose a novel TFAA module and PCA-FPN to enhance the feature from sequential multimedia data. We also provide a new dataset called TVIL for TFL in a novel scenario which has been released for academic use. The experimental results show the effectiveness of the proposed framework. Especially concerning the LAV-DF dataset, compared to the previous state-of-the-art method BA-TFD~\cite{DBLP:journals/corr/abs-2204-06228}, our approach has shown significant performance improvements. Specifically, the AP has increased from $76.90\%$ to $98.83\%$ at tIOU 0.5, and from $0.25\%$ to $37.61\%$ at tIOU 0.95. In the future, we will conduct further research to spatial localization on top of temporal localization to enhance the practicality of the model.

\begin{acks}
This work is supported by the National Natural Science Foundation of China (NSFC) under Grants 62272331 and 61972269, Sichuan Science and Technology Program under Grant 2022YFG0320.
\end{acks}

\bibliographystyle{ACM-Reference-Format}
\balance
\bibliography{ref}


\begin{thebibliography}{66}


\ifx \showCODEN    \undefined \def \showCODEN     #1{\unskip}     \fi
\ifx \showDOI      \undefined \def \showDOI       #1{#1}\fi
\ifx \showISBNx    \undefined \def \showISBNx     #1{\unskip}     \fi
\ifx \showISBNxiii \undefined \def \showISBNxiii  #1{\unskip}     \fi
\ifx \showISSN     \undefined \def \showISSN      #1{\unskip}     \fi
\ifx \showLCCN     \undefined \def \showLCCN      #1{\unskip}     \fi
\ifx \shownote     \undefined \def \shownote      #1{#1}          \fi
\ifx \showarticletitle \undefined \def \showarticletitle #1{#1}   \fi
\ifx \showURL      \undefined \def \showURL       {\relax}        \fi
\providecommand\bibfield[2]{#2}
\providecommand\bibinfo[2]{#2}
\providecommand\natexlab[1]{#1}
\providecommand\showeprint[2][]{arXiv:#2}

\bibitem[Aldausari et~al\mbox{.}(2023)]%
        {DBLP:journals/csur/AldausariSMM23}
\bibfield{author}{\bibinfo{person}{Nuha Aldausari}, \bibinfo{person}{Arcot
  Sowmya}, \bibinfo{person}{Nadine Marcus}, {and} \bibinfo{person}{Gelareh
  Mohammadi}.} \bibinfo{year}{2023}\natexlab{}.
\newblock \showarticletitle{Video Generative Adversarial Networks: {A} Review}.
\newblock \bibinfo{journal}{\emph{{ACM} Comput. Surv.}} \bibinfo{volume}{55},
  \bibinfo{number}{2} (\bibinfo{year}{2023}), \bibinfo{pages}{30:1--30:25}.
\newblock
\urldef\tempurl%
\url{https://doi.org/10.1145/3487891}
\showDOI{\tempurl}


\bibitem[Ba et~al\mbox{.}(2016)]%
        {DBLP:journals/corr/BaKH16}
\bibfield{author}{\bibinfo{person}{Jimmy~Lei Ba}, \bibinfo{person}{Jamie~Ryan
  Kiros}, {and} \bibinfo{person}{Geoffrey~E. Hinton}.}
  \bibinfo{year}{2016}\natexlab{}.
\newblock \bibinfo{title}{Layer Normalization}.
\newblock
\newblock
\showeprint[arxiv]{1607.06450}


\bibitem[Bagchi. et~al\mbox{.}(2022)]%
        {DBLP:conf/visapp/BagchiMFS22}
\bibfield{author}{\bibinfo{person}{Anurag Bagchi.}, \bibinfo{person}{Jazib
  Mahmood.}, \bibinfo{person}{Dolton Fernandes.}, {and}
  \bibinfo{person}{Ravi~Kiran Sarvadevabhatla.}}
  \bibinfo{year}{2022}\natexlab{}.
\newblock \showarticletitle{Hear Me out: Fusional Approaches for Audio
  Augmented Temporal Action Localization}. In
  \bibinfo{booktitle}{\emph{Proceedings of the 17th International Joint
  Conference on Computer Vision, Imaging and Computer Graphics Theory and
  Applications}}. \bibinfo{publisher}{SciTePress}, \bibinfo{pages}{144--154}.
\newblock
\urldef\tempurl%
\url{https://doi.org/10.5220/0010832700003124}
\showDOI{\tempurl}


\bibitem[Bodla et~al\mbox{.}(2017)]%
        {DBLP:conf/iccv/BodlaSCD17}
\bibfield{author}{\bibinfo{person}{Navaneeth Bodla}, \bibinfo{person}{Bharat
  Singh}, \bibinfo{person}{Rama Chellappa}, {and} \bibinfo{person}{Larry~S.
  Davis}.} \bibinfo{year}{2017}\natexlab{}.
\newblock \showarticletitle{Soft-NMS - Improving Object Detection with One Line
  of Code}. In \bibinfo{booktitle}{\emph{Proceedings of the {IEEE}
  International Conference on Computer Vision}}. \bibinfo{publisher}{{IEEE}},
  \bibinfo{pages}{5562--5570}.
\newblock
\urldef\tempurl%
\url{https://doi.org/10.1109/ICCV.2017.593}
\showDOI{\tempurl}


\bibitem[Cai et~al\mbox{.}(2022a)]%
        {DBLP:conf/mm/CaiLTYT22}
\bibfield{author}{\bibinfo{person}{Jiayin Cai}, \bibinfo{person}{Changlin Li},
  \bibinfo{person}{Xin Tao}, \bibinfo{person}{Chun Yuan}, {and}
  \bibinfo{person}{Yu{-}Wing Tai}.} \bibinfo{year}{2022}\natexlab{a}.
\newblock \showarticletitle{DeViT: Deformed Vision Transformers in Video
  Inpainting}. In \bibinfo{booktitle}{\emph{Proceedings of the 30th {ACM}
  International Conference on Multimedia}}. \bibinfo{publisher}{{ACM}},
  \bibinfo{pages}{779--789}.
\newblock
\urldef\tempurl%
\url{https://doi.org/10.1145/3503161.3548395}
\showDOI{\tempurl}


\bibitem[Cai et~al\mbox{.}(2022b)]%
        {DBLP:journals/corr/abs-2204-06228}
\bibfield{author}{\bibinfo{person}{Zhixi Cai}, \bibinfo{person}{Kalin
  Stefanov}, \bibinfo{person}{Abhinav Dhall}, {and} \bibinfo{person}{Munawar
  Hayat}.} \bibinfo{year}{2022}\natexlab{b}.
\newblock \showarticletitle{Do You Really Mean That? Content Driven
  Audio-Visual Deepfake Dataset and Multimodal Method for Temporal Forgery
  Localization}. In \bibinfo{booktitle}{\emph{2022 International Conference on
  Digital Image Computing: Techniques and Applications (DICTA)}}.
  \bibinfo{publisher}{{IEEE}}, \bibinfo{pages}{1--10}.
\newblock
\urldef\tempurl%
\url{https://doi.org/10.1109/DICTA56598.2022.10034605}
\showDOI{\tempurl}


\bibitem[Cao et~al\mbox{.}(2022)]%
        {DBLP:conf/cvpr/Cao0YCDY22}
\bibfield{author}{\bibinfo{person}{Junyi Cao}, \bibinfo{person}{Chao Ma},
  \bibinfo{person}{Taiping Yao}, \bibinfo{person}{Shen Chen},
  \bibinfo{person}{Shouhong Ding}, {and} \bibinfo{person}{Xiaokang Yang}.}
  \bibinfo{year}{2022}\natexlab{}.
\newblock \showarticletitle{End-to-End Reconstruction-Classification Learning
  for Face Forgery Detection}. In \bibinfo{booktitle}{\emph{Proceedings of the
  IEEE Conference on Computer Vision and Pattern Recognition}}.
  \bibinfo{publisher}{{IEEE}}, \bibinfo{pages}{4103--4112}.
\newblock
\urldef\tempurl%
\url{https://doi.org/10.1109/CVPR52688.2022.00408}
\showDOI{\tempurl}


\bibitem[Chen et~al\mbox{.}(2022)]%
        {DBLP:conf/aaai/ChenZWL22}
\bibfield{author}{\bibinfo{person}{Guo Chen}, \bibinfo{person}{Yin{-}Dong
  Zheng}, \bibinfo{person}{Limin Wang}, {and} \bibinfo{person}{Tong Lu}.}
  \bibinfo{year}{2022}\natexlab{}.
\newblock \showarticletitle{{DCAN:} Improving Temporal Action Detection via
  Dual Context Aggregation}. In \bibinfo{booktitle}{\emph{Proceedings of the
  {AAAI} Conference on Artificial Intelligence}}. \bibinfo{publisher}{{AAAI}
  Press}, \bibinfo{pages}{248--257}.
\newblock


\bibitem[Chen et~al\mbox{.}(2023)]%
        {DBLP:conf/aaai/Wang_Chow_2023}
\bibfield{author}{\bibinfo{person}{Guo Chen}, \bibinfo{person}{Yin{-}Dong
  Zheng}, \bibinfo{person}{Limin Wang}, {and} \bibinfo{person}{Tong Lu}.}
  \bibinfo{year}{2023}\natexlab{}.
\newblock \showarticletitle{Noise Based Deepfake Detection via Multi-Head
  Relative-Interaction}. In \bibinfo{booktitle}{\emph{Proceedings of the {AAAI}
  Conference on Artificial Intelligence}}. \bibinfo{publisher}{{AAAI} Press},
  \bibinfo{pages}{14548--14556}.
\newblock


\bibitem[Cheng and Schwing(2022)]%
        {DBLP:conf/eccv/ChengS22}
\bibfield{author}{\bibinfo{person}{Ho~Kei Cheng} {and}
  \bibinfo{person}{Alexander~G. Schwing}.} \bibinfo{year}{2022}\natexlab{}.
\newblock \showarticletitle{XMem: Long-Term Video Object Segmentation with an
  Atkinson-Shiffrin Memory Model}. In \bibinfo{booktitle}{\emph{European
  Conference on Computer Vision}}. \bibinfo{publisher}{Springer},
  \bibinfo{pages}{640--658}.
\newblock
\urldef\tempurl%
\url{https://doi.org/10.1007/978-3-031-19815-1\_37}
\showDOI{\tempurl}


\bibitem[Chugh et~al\mbox{.}(2020)]%
        {DBLP:conf/mm/ChughGDS20}
\bibfield{author}{\bibinfo{person}{Komal Chugh}, \bibinfo{person}{Parul Gupta},
  \bibinfo{person}{Abhinav Dhall}, {and} \bibinfo{person}{Ramanathan
  Subramanian}.} \bibinfo{year}{2020}\natexlab{}.
\newblock \showarticletitle{Not made for each other- Audio-Visual
  Dissonance-based Deepfake Detection and Localization}. In
  \bibinfo{booktitle}{\emph{Proceedings of the 28th {ACM} international
  conference on Multimedia}}. \bibinfo{publisher}{{ACM}},
  \bibinfo{pages}{439--447}.
\newblock
\urldef\tempurl%
\url{https://doi.org/10.1145/3394171.3413700}
\showDOI{\tempurl}


\bibitem[Coccomini et~al\mbox{.}(2022)]%
        {DBLP:conf/iciap/CoccominiMGF22}
\bibfield{author}{\bibinfo{person}{Davide Coccomini}, \bibinfo{person}{Nicola
  Messina}, \bibinfo{person}{Claudio Gennaro}, {and} \bibinfo{person}{Fabrizio
  Falchi}.} \bibinfo{year}{2022}\natexlab{}.
\newblock \showarticletitle{Combining EfficientNet and Vision Transformers for
  Video Deepfake Detection}. In \bibinfo{booktitle}{\emph{Proceedings of
  International Conference on Image Analysis and Processing}}.
  \bibinfo{publisher}{Springer}, \bibinfo{pages}{219--229}.
\newblock
\urldef\tempurl%
\url{https://doi.org/10.1007/978-3-031-06433-3\_19}
\showDOI{\tempurl}


\bibitem[Deng et~al\mbox{.}(2022)]%
        {DBLP:conf/mm/DengMY0D22}
\bibfield{author}{\bibinfo{person}{Jiacheng Deng}, \bibinfo{person}{Terui Mao},
  \bibinfo{person}{Diqun Yan}, \bibinfo{person}{Li Dong}, {and}
  \bibinfo{person}{Mingyu Dong}.} \bibinfo{year}{2022}\natexlab{}.
\newblock \showarticletitle{Detection of Synthetic Speech Based on Spectrum
  Defects}. In \bibinfo{booktitle}{\emph{Proceedings of the 1st International
  Workshop on Deepfake Detection for Audio Multimedia}}.
  \bibinfo{publisher}{{ACM}}, \bibinfo{pages}{3--8}.
\newblock
\urldef\tempurl%
\url{https://doi.org/10.1145/3552466.3556529}
\showDOI{\tempurl}


\bibitem[Dolhansky et~al\mbox{.}(2020)]%
        {DBLP:journals/corr/abs-2006-07397}
\bibfield{author}{\bibinfo{person}{Brian Dolhansky}, \bibinfo{person}{Joanna
  Bitton}, \bibinfo{person}{Ben Pflaum}, \bibinfo{person}{Jikuo Lu},
  \bibinfo{person}{Russ Howes}, \bibinfo{person}{Menglin Wang}, {and}
  \bibinfo{person}{Cristian~Canton Ferrer}.} \bibinfo{year}{2020}\natexlab{}.
\newblock \bibinfo{title}{The DeepFake Detection Challenge (DFDC) Dataset}.
\newblock
\newblock
\showeprint[arxiv]{2006.07397}


\bibitem[Dolhansky et~al\mbox{.}(2019)]%
        {DBLP:journals/corr/abs-1910-08854}
\bibfield{author}{\bibinfo{person}{Brian Dolhansky}, \bibinfo{person}{Russ
  Howes}, \bibinfo{person}{Ben Pflaum}, \bibinfo{person}{Nicole Baram}, {and}
  \bibinfo{person}{Cristian~Canton Ferrer}.} \bibinfo{year}{2019}\natexlab{}.
\newblock \bibinfo{title}{The Deepfake Detection Challenge (DFDC) Preview
  Dataset}.
\newblock
\newblock
\showeprint[arxiv]{1910.08854}


\bibitem[Feichtenhofer et~al\mbox{.}(2019)]%
        {DBLP:conf/iccv/Feichtenhofer0M19}
\bibfield{author}{\bibinfo{person}{Christoph Feichtenhofer},
  \bibinfo{person}{Haoqi Fan}, \bibinfo{person}{Jitendra Malik}, {and}
  \bibinfo{person}{Kaiming He}.} \bibinfo{year}{2019}\natexlab{}.
\newblock \showarticletitle{SlowFast Networks for Video Recognition}. In
  \bibinfo{booktitle}{\emph{{IEEE} International Conference on Computer
  Vision}}. \bibinfo{publisher}{{IEEE}}, \bibinfo{pages}{6201--6210}.
\newblock
\urldef\tempurl%
\url{https://doi.org/10.1109/ICCV.2019.00630}
\showDOI{\tempurl}


\bibitem[Frank and Sch\"{o}nherr(2021)]%
        {DBLP:conf/nips/FrankS21}
\bibfield{author}{\bibinfo{person}{Joel Frank} {and} \bibinfo{person}{Lea
  Sch\"{o}nherr}.} \bibinfo{year}{2021}\natexlab{}.
\newblock \showarticletitle{WaveFake: A Data Set to Facilitate Audio Deepfake
  Detection}. In \bibinfo{booktitle}{\emph{Proceedings of the Neural
  Information Processing Systems Track on Datasets and Benchmarks}},
  Vol.~\bibinfo{volume}{1}. \bibinfo{publisher}{Curran Associates, Inc.}
\newblock


\bibitem[Gemmeke et~al\mbox{.}(2017)]%
        {DBLP:conf/icassp/GemmekeEFJLMPR17}
\bibfield{author}{\bibinfo{person}{Jort~F. Gemmeke}, \bibinfo{person}{Daniel
  P.~W. Ellis}, \bibinfo{person}{Dylan Freedman}, \bibinfo{person}{Aren
  Jansen}, \bibinfo{person}{Wade Lawrence}, \bibinfo{person}{R.~Channing
  Moore}, \bibinfo{person}{Manoj Plakal}, {and} \bibinfo{person}{Marvin
  Ritter}.} \bibinfo{year}{2017}\natexlab{}.
\newblock \showarticletitle{Audio Set: An ontology and human-labeled dataset
  for audio events}. In \bibinfo{booktitle}{\emph{{IEEE} International
  Conference on Acoustics, Speech and Signal Processing}}.
  \bibinfo{publisher}{{IEEE}}, \bibinfo{pages}{776--780}.
\newblock
\urldef\tempurl%
\url{https://doi.org/10.1109/ICASSP.2017.7952261}
\showDOI{\tempurl}


\bibitem[Guo et~al\mbox{.}(2022)]%
        {DBLP:conf/mm/GuoTDDK22}
\bibfield{author}{\bibinfo{person}{Jiwei Guo}, \bibinfo{person}{Jiajia Tang},
  \bibinfo{person}{Weichen Dai}, \bibinfo{person}{Yu Ding}, {and}
  \bibinfo{person}{Wanzeng Kong}.} \bibinfo{year}{2022}\natexlab{}.
\newblock \showarticletitle{Dynamically Adjust Word Representations Using
  Unaligned Multimodal Information}. In \bibinfo{booktitle}{\emph{Proceedings
  of the 30th {ACM} international conference on Multimedia}}.
  \bibinfo{publisher}{{ACM}}, \bibinfo{pages}{3394--3402}.
\newblock
\urldef\tempurl%
\url{https://doi.org/10.1145/3503161.3548137}
\showDOI{\tempurl}


\bibitem[Han et~al\mbox{.}(2021)]%
        {DBLP:journals/tbbis/HanHZLC21}
\bibfield{author}{\bibinfo{person}{Bing Han}, \bibinfo{person}{Xiaoguang Han},
  \bibinfo{person}{Hua Zhang}, \bibinfo{person}{Jingzhi Li}, {and}
  \bibinfo{person}{Xiaochun Cao}.} \bibinfo{year}{2021}\natexlab{}.
\newblock \showarticletitle{Fighting Fake News: Two Stream Network for Deepfake
  Detection via Learnable {SRM}}.
\newblock \bibinfo{journal}{\emph{{IEEE} Trans. Biom. Behav. Identity Sci.}}
  \bibinfo{volume}{3}, \bibinfo{number}{3} (\bibinfo{year}{2021}),
  \bibinfo{pages}{320--331}.
\newblock
\urldef\tempurl%
\url{https://doi.org/10.1109/TBIOM.2021.3065735}
\showDOI{\tempurl}


\bibitem[He et~al\mbox{.}(2021)]%
        {DBLP:conf/cvpr/HeGCZYSSS021}
\bibfield{author}{\bibinfo{person}{Yinan He}, \bibinfo{person}{Bei Gan},
  \bibinfo{person}{Siyu Chen}, \bibinfo{person}{Yichun Zhou},
  \bibinfo{person}{Guojun Yin}, \bibinfo{person}{Luchuan Song},
  \bibinfo{person}{Lu Sheng}, \bibinfo{person}{Jing Shao}, {and}
  \bibinfo{person}{Ziwei Liu}.} \bibinfo{year}{2021}\natexlab{}.
\newblock \showarticletitle{ForgeryNet: {A} Versatile Benchmark for
  Comprehensive Forgery Analysis}. In \bibinfo{booktitle}{\emph{Proceedings of
  the IEEE Conference on Computer Vision and Pattern Recognition}}.
  \bibinfo{publisher}{{IEEE}}, \bibinfo{pages}{4360--4369}.
\newblock
\urldef\tempurl%
\url{https://doi.org/10.1109/CVPR46437.2021.00434}
\showDOI{\tempurl}


\bibitem[Hegde et~al\mbox{.}(2022)]%
        {DBLP:conf/mm/HegdePMNJ22}
\bibfield{author}{\bibinfo{person}{Sindhu~B. Hegde}, \bibinfo{person}{K.~R.
  Prajwal}, \bibinfo{person}{Rudrabha Mukhopadhyay}, \bibinfo{person}{Vinay~P.
  Namboodiri}, {and} \bibinfo{person}{C.~V. Jawahar}.}
  \bibinfo{year}{2022}\natexlab{}.
\newblock \showarticletitle{Lip-to-Speech Synthesis for Arbitrary Speakers in
  the Wild}. In \bibinfo{booktitle}{\emph{Proceedings of the 30th {ACM}
  International Conference on Multimedia}}. \bibinfo{publisher}{{ACM}},
  \bibinfo{pages}{6250--6258}.
\newblock
\urldef\tempurl%
\url{https://doi.org/10.1145/3503161.3548081}
\showDOI{\tempurl}


\bibitem[Heilbron et~al\mbox{.}(2015)]%
        {DBLP:conf/cvpr/HeilbronEGN15}
\bibfield{author}{\bibinfo{person}{Fabian~Caba Heilbron},
  \bibinfo{person}{Victor Escorcia}, \bibinfo{person}{Bernard Ghanem}, {and}
  \bibinfo{person}{Juan~Carlos Niebles}.} \bibinfo{year}{2015}\natexlab{}.
\newblock \showarticletitle{ActivityNet: {A} large-scale video benchmark for
  human activity understanding}. In \bibinfo{booktitle}{\emph{Proceedings of
  the IEEE Conference on Computer Vision and Pattern Recognition}}.
  \bibinfo{publisher}{{IEEE}}, \bibinfo{pages}{961--970}.
\newblock
\urldef\tempurl%
\url{https://doi.org/10.1109/CVPR.2015.7298698}
\showDOI{\tempurl}


\bibitem[Hershey et~al\mbox{.}(2017)]%
        {DBLP:conf/icassp/HersheyCEGJMPPS17}
\bibfield{author}{\bibinfo{person}{Shawn Hershey}, \bibinfo{person}{Sourish
  Chaudhuri}, \bibinfo{person}{Daniel P.~W. Ellis}, \bibinfo{person}{Jort~F.
  Gemmeke}, \bibinfo{person}{Aren Jansen}, \bibinfo{person}{R.~Channing Moore},
  \bibinfo{person}{Manoj Plakal}, \bibinfo{person}{Devin Platt},
  \bibinfo{person}{Rif~A. Saurous}, \bibinfo{person}{Bryan Seybold},
  \bibinfo{person}{Malcolm Slaney}, \bibinfo{person}{Ron~J. Weiss}, {and}
  \bibinfo{person}{Kevin~W. Wilson}.} \bibinfo{year}{2017}\natexlab{}.
\newblock \showarticletitle{{CNN} architectures for large-scale audio
  classification}. In \bibinfo{booktitle}{\emph{{IEEE} International Conference
  on Acoustics, Speech and Signal Processing}}. \bibinfo{publisher}{{IEEE}},
  \bibinfo{pages}{131--135}.
\newblock
\urldef\tempurl%
\url{https://doi.org/10.1109/ICASSP.2017.7952132}
\showDOI{\tempurl}


\bibitem[Hu et~al\mbox{.}(2022)]%
        {DBLP:journals/tcsv/HuLWQ22}
\bibfield{author}{\bibinfo{person}{Juan Hu}, \bibinfo{person}{Xin Liao},
  \bibinfo{person}{Wei Wang}, {and} \bibinfo{person}{Zheng Qin}.}
  \bibinfo{year}{2022}\natexlab{}.
\newblock \showarticletitle{Detecting Compressed Deepfake Videos in Social
  Networks Using Frame-Temporality Two-Stream Convolutional Network}.
\newblock \bibinfo{journal}{\emph{{IEEE} Trans. Circuits Syst. Video Technol.}}
  \bibinfo{volume}{32}, \bibinfo{number}{3} (\bibinfo{year}{2022}),
  \bibinfo{pages}{1089--1102}.
\newblock
\urldef\tempurl%
\url{https://doi.org/10.1109/CVPRW.2017.229}
\showDOI{\tempurl}


\bibitem[Hu and Singh(2021)]%
        {DBLP:conf/iccv/HuS21}
\bibfield{author}{\bibinfo{person}{Ronghang Hu} {and}
  \bibinfo{person}{Amanpreet Singh}.} \bibinfo{year}{2021}\natexlab{}.
\newblock \showarticletitle{UniT: Multimodal Multitask Learning with a Unified
  Transformer}. In \bibinfo{booktitle}{\emph{Proceedings of the {IEEE}
  International Conference on Computer Vision}}. \bibinfo{publisher}{{IEEE}},
  \bibinfo{pages}{1419--1429}.
\newblock
\urldef\tempurl%
\url{https://doi.org/10.1109/ICCV48922.2021.00147}
\showDOI{\tempurl}


\bibitem[Khalid et~al\mbox{.}(2021)]%
        {DBLP:conf/nips/KhalidTKW21}
\bibfield{author}{\bibinfo{person}{Hasam Khalid}, \bibinfo{person}{Shahroz
  Tariq}, \bibinfo{person}{Minha Kim}, {and} \bibinfo{person}{Simon Woo}.}
  \bibinfo{year}{2021}\natexlab{}.
\newblock \showarticletitle{FakeAVCeleb: A Novel Audio-Video Multimodal
  Deepfake Dataset}. In \bibinfo{booktitle}{\emph{Proceedings of the Neural
  Information Processing Systems Track on Datasets and Benchmarks}},
  Vol.~\bibinfo{volume}{1}. \bibinfo{publisher}{Curran Associates, Inc.}
\newblock


\bibitem[Kwak et~al\mbox{.}(2022)]%
        {DBLP:conf/mm/KwakCYLHO22}
\bibfield{author}{\bibinfo{person}{Il{-}Youp Kwak}, \bibinfo{person}{Sunmook
  Choi}, \bibinfo{person}{Jonghoon Yang}, \bibinfo{person}{Yerin Lee},
  \bibinfo{person}{Soyul Han}, {and} \bibinfo{person}{Seungsang Oh}.}
  \bibinfo{year}{2022}\natexlab{}.
\newblock \showarticletitle{Low-quality Fake Audio Detection through Frequency
  Feature Masking}. In \bibinfo{booktitle}{\emph{Proceedings of the 1st
  International Workshop on Deepfake Detection for Audio Multimedia}}.
  \bibinfo{publisher}{{ACM}}, \bibinfo{pages}{9--17}.
\newblock
\urldef\tempurl%
\url{https://doi.org/10.1145/3552466.3556533}
\showDOI{\tempurl}


\bibitem[Li et~al\mbox{.}(2020)]%
        {DBLP:conf/cvpr/LiBZYCWG20}
\bibfield{author}{\bibinfo{person}{Lingzhi Li}, \bibinfo{person}{Jianmin Bao},
  \bibinfo{person}{Ting Zhang}, \bibinfo{person}{Hao Yang},
  \bibinfo{person}{Dong Chen}, \bibinfo{person}{Fang Wen}, {and}
  \bibinfo{person}{Baining Guo}.} \bibinfo{year}{2020}\natexlab{}.
\newblock \showarticletitle{Face X-Ray for More General Face Forgery
  Detection}. In \bibinfo{booktitle}{\emph{Proceedings of the IEEE Conference
  on Computer Vision and Pattern Recognition}}. \bibinfo{publisher}{{IEEE}},
  \bibinfo{pages}{5000--5009}.
\newblock
\urldef\tempurl%
\url{https://doi.org/10.1109/CVPR42600.2020.00505}
\showDOI{\tempurl}


\bibitem[Li et~al\mbox{.}(2022a)]%
        {DBLP:conf/mm/LiH0SYY22}
\bibfield{author}{\bibinfo{person}{Yudong Li}, \bibinfo{person}{Xianxu Hou},
  \bibinfo{person}{Zhe Zhao}, \bibinfo{person}{Linlin Shen},
  \bibinfo{person}{Xuefeng Yang}, {and} \bibinfo{person}{Kimmo Yan}.}
  \bibinfo{year}{2022}\natexlab{a}.
\newblock \showarticletitle{Talk2Face: {A} Unified Sequence-based Framework for
  Diverse Face Generation and Analysis Tasks}. In
  \bibinfo{booktitle}{\emph{Proceedings of the 30th {ACM} International
  Conference on Multimedia}}. \bibinfo{publisher}{{ACM}},
  \bibinfo{pages}{4594--4604}.
\newblock
\urldef\tempurl%
\url{https://doi.org/10.1145/3503161.3548205}
\showDOI{\tempurl}


\bibitem[Li et~al\mbox{.}(2021)]%
        {DBLP:journals/taslp/LiLZFZ21}
\bibfield{author}{\bibinfo{person}{Zekang Li}, \bibinfo{person}{Zongjia Li},
  \bibinfo{person}{Jinchao Zhang}, \bibinfo{person}{Yang Feng}, {and}
  \bibinfo{person}{Jie Zhou}.} \bibinfo{year}{2021}\natexlab{}.
\newblock \showarticletitle{Bridging Text and Video: {A} Universal Multimodal
  Transformer for Audio-Visual Scene-Aware Dialog}.
\newblock \bibinfo{journal}{\emph{{IEEE} {ACM} Trans. Audio Speech Lang.
  Process.}}  \bibinfo{volume}{29} (\bibinfo{year}{2021}),
  \bibinfo{pages}{2476--2483}.
\newblock
\urldef\tempurl%
\url{https://doi.org/10.1109/TASLP.2021.3065823}
\showDOI{\tempurl}


\bibitem[Li et~al\mbox{.}(2022b)]%
        {DBLP:conf/cvpr/0031LQGC22}
\bibfield{author}{\bibinfo{person}{Zhen Li}, \bibinfo{person}{Chengze Lu},
  \bibinfo{person}{Jianhua Qin}, \bibinfo{person}{Chun{-}Le Guo}, {and}
  \bibinfo{person}{Ming{-}Ming Cheng}.} \bibinfo{year}{2022}\natexlab{b}.
\newblock \showarticletitle{Towards An End-to-End Framework for Flow-Guided
  Video Inpainting}. In \bibinfo{booktitle}{\emph{Proceedings of the IEEE
  Conference on Computer Vision and Pattern Recognition}}.
  \bibinfo{publisher}{{IEEE}}, \bibinfo{pages}{17541--17550}.
\newblock
\urldef\tempurl%
\url{https://doi.org/10.1109/CVPR52688.2022.01704}
\showDOI{\tempurl}


\bibitem[Lin et~al\mbox{.}(2021)]%
        {DBLP:conf/cvpr/Lin0LWTWLHF21}
\bibfield{author}{\bibinfo{person}{Chuming Lin}, \bibinfo{person}{Chengming
  Xu}, \bibinfo{person}{Donghao Luo}, \bibinfo{person}{Yabiao Wang},
  \bibinfo{person}{Ying Tai}, \bibinfo{person}{Chengjie Wang},
  \bibinfo{person}{Jilin Li}, \bibinfo{person}{Feiyue Huang}, {and}
  \bibinfo{person}{Yanwei Fu}.} \bibinfo{year}{2021}\natexlab{}.
\newblock \showarticletitle{Learning Salient Boundary Feature for Anchor-free
  Temporal Action Localization}. In \bibinfo{booktitle}{\emph{Proceedings of
  the IEEE Conference on Computer Vision and Pattern Recognition}}.
  \bibinfo{publisher}{{IEEE}}, \bibinfo{pages}{3320--3329}.
\newblock
\urldef\tempurl%
\url{https://doi.org/10.1109/CVPR46437.2021.00333}
\showDOI{\tempurl}


\bibitem[Lin et~al\mbox{.}(2017a)]%
        {DBLP:conf/cvpr/LinDGHHB17}
\bibfield{author}{\bibinfo{person}{Tsung{-}Yi Lin}, \bibinfo{person}{Piotr
  Doll{\'{a}}r}, \bibinfo{person}{Ross~B. Girshick}, \bibinfo{person}{Kaiming
  He}, \bibinfo{person}{Bharath Hariharan}, {and} \bibinfo{person}{Serge~J.
  Belongie}.} \bibinfo{year}{2017}\natexlab{a}.
\newblock \showarticletitle{Feature Pyramid Networks for Object Detection}. In
  \bibinfo{booktitle}{\emph{Proceedings of the IEEE Conference on Computer
  Vision and Pattern Recognition}}. \bibinfo{publisher}{{IEEE}},
  \bibinfo{pages}{936--944}.
\newblock
\urldef\tempurl%
\url{https://doi.org/10.1109/CVPR.2017.106}
\showDOI{\tempurl}


\bibitem[Lin et~al\mbox{.}(2017b)]%
        {DBLP:conf/iccv/LinGGHD17}
\bibfield{author}{\bibinfo{person}{Tsung{-}Yi Lin}, \bibinfo{person}{Priya
  Goyal}, \bibinfo{person}{Ross~B. Girshick}, \bibinfo{person}{Kaiming He},
  {and} \bibinfo{person}{Piotr Doll{\'{a}}r}.}
  \bibinfo{year}{2017}\natexlab{b}.
\newblock \showarticletitle{Focal Loss for Dense Object Detection}. In
  \bibinfo{booktitle}{\emph{Proceedings of the {IEEE} International Conference
  on Computer Vision}}. \bibinfo{publisher}{{IEEE}},
  \bibinfo{pages}{2999--3007}.
\newblock
\urldef\tempurl%
\url{https://doi.org/10.1109/ICCV.2017.324}
\showDOI{\tempurl}


\bibitem[Lin et~al\mbox{.}(2019)]%
        {DBLP:conf/iccv/LinLLDW19}
\bibfield{author}{\bibinfo{person}{Tianwei Lin}, \bibinfo{person}{Xiao Liu},
  \bibinfo{person}{Xin Li}, \bibinfo{person}{Errui Ding}, {and}
  \bibinfo{person}{Shilei Wen}.} \bibinfo{year}{2019}\natexlab{}.
\newblock \showarticletitle{{BMN:} Boundary-Matching Network for Temporal
  Action Proposal Generation}. In \bibinfo{booktitle}{\emph{Proceedings of the
  {IEEE} International Conference on Computer Vision}}.
  \bibinfo{publisher}{{IEEE}}, \bibinfo{pages}{3888--3897}.
\newblock
\urldef\tempurl%
\url{https://doi.org/10.1109/ICCV.2019.00399}
\showDOI{\tempurl}


\bibitem[Liu et~al\mbox{.}(2021)]%
        {DBLP:conf/iccv/0019DHSLS0D021}
\bibfield{author}{\bibinfo{person}{Rui Liu}, \bibinfo{person}{Hanming Deng},
  \bibinfo{person}{Yangyi Huang}, \bibinfo{person}{Xiaoyu Shi},
  \bibinfo{person}{Lewei Lu}, \bibinfo{person}{Wenxiu Sun},
  \bibinfo{person}{Xiaogang Wang}, \bibinfo{person}{Jifeng Dai}, {and}
  \bibinfo{person}{Hongsheng Li}.} \bibinfo{year}{2021}\natexlab{}.
\newblock \showarticletitle{FuseFormer: Fusing Fine-Grained Information in
  Transformers for Video Inpainting}. In \bibinfo{booktitle}{\emph{Proceedings
  of the {IEEE} International Conference on Computer Vision}}.
  \bibinfo{publisher}{{IEEE}}, \bibinfo{pages}{14020--14029}.
\newblock
\urldef\tempurl%
\url{https://doi.org/10.1109/ICCV48922.2021.01378}
\showDOI{\tempurl}


\bibitem[Mittal et~al\mbox{.}(2023)]%
        {DBLP:conf/wacv/MittalSSCM23}
\bibfield{author}{\bibinfo{person}{Trisha Mittal}, \bibinfo{person}{Ritwik
  Sinha}, \bibinfo{person}{Viswanathan Swaminathan}, \bibinfo{person}{John~P.
  Collomosse}, {and} \bibinfo{person}{Dinesh Manocha}.}
  \bibinfo{year}{2023}\natexlab{}.
\newblock \showarticletitle{Video Manipulations Beyond Faces: {A} Dataset with
  Human-Machine Analysis}. In \bibinfo{booktitle}{\emph{{IEEE/CVF} Winter
  Conference on Applications of Computer Vision Workshops}}.
  \bibinfo{publisher}{{IEEE}}, \bibinfo{pages}{643--652}.
\newblock
\urldef\tempurl%
\url{https://doi.org/10.1109/WACVW58289.2023.00071}
\showDOI{\tempurl}


\bibitem[Nag et~al\mbox{.}(2022)]%
        {DBLP:conf/eccv/NagZSX22}
\bibfield{author}{\bibinfo{person}{Sauradip Nag}, \bibinfo{person}{Xiatian
  Zhu}, \bibinfo{person}{Yi{-}Zhe Song}, {and} \bibinfo{person}{Tao Xiang}.}
  \bibinfo{year}{2022}\natexlab{}.
\newblock \showarticletitle{Proposal-Free Temporal Action Detection via Global
  Segmentation Mask Learning}. In \bibinfo{booktitle}{\emph{European Conference
  on Computer Vision}}. \bibinfo{publisher}{Springer},
  \bibinfo{pages}{645--662}.
\newblock
\urldef\tempurl%
\url{https://doi.org/10.1007/978-3-031-20062-5\_37}
\showDOI{\tempurl}


\bibitem[Nawhal and Mori(2021)]%
        {DBLP:journals/corr/abs-2101-08540}
\bibfield{author}{\bibinfo{person}{Megha Nawhal} {and} \bibinfo{person}{Greg
  Mori}.} \bibinfo{year}{2021}\natexlab{}.
\newblock \bibinfo{title}{Activity Graph Transformer for Temporal Action
  Localization}.
\newblock
\newblock
\showeprint[arxiv]{2101.08540}


\bibitem[Nguyen et~al\mbox{.}(2019)]%
        {DBLP:conf/icassp/NguyenYE19}
\bibfield{author}{\bibinfo{person}{Huy~H. Nguyen}, \bibinfo{person}{Junichi
  Yamagishi}, {and} \bibinfo{person}{Isao Echizen}.}
  \bibinfo{year}{2019}\natexlab{}.
\newblock \showarticletitle{Capsule-forensics: Using Capsule Networks to Detect
  Forged Images and Videos}. In \bibinfo{booktitle}{\emph{{IEEE} International
  Conference on Acoustics, Speech and Signal Processing}}.
  \bibinfo{publisher}{{IEEE}}, \bibinfo{pages}{2307--2311}.
\newblock
\urldef\tempurl%
\url{https://doi.org/10.1109/ICASSP.2019.8682602}
\showDOI{\tempurl}


\bibitem[Niizumi et~al\mbox{.}(2021)]%
        {DBLP:conf/ijcnn/NiizumiTOHK21}
\bibfield{author}{\bibinfo{person}{Daisuke Niizumi}, \bibinfo{person}{Daiki
  Takeuchi}, \bibinfo{person}{Yasunori Ohishi}, \bibinfo{person}{Noboru
  Harada}, {and} \bibinfo{person}{Kunio Kashino}.}
  \bibinfo{year}{2021}\natexlab{}.
\newblock \showarticletitle{{BYOL} for Audio: Self-Supervised Learning for
  General-Purpose Audio Representation}. In
  \bibinfo{booktitle}{\emph{International Joint Conference on Neural
  Networks}}. \bibinfo{publisher}{{IEEE}}, \bibinfo{pages}{1--8}.
\newblock
\urldef\tempurl%
\url{https://doi.org/10.1109/IJCNN52387.2021.9534474}
\showDOI{\tempurl}


\bibitem[Perov et~al\mbox{.}(2021)]%
        {DBLP:journals/corr/abs-2005-05535}
\bibfield{author}{\bibinfo{person}{Ivan Perov}, \bibinfo{person}{Daiheng Gao},
  \bibinfo{person}{Nikolay Chervoniy}, \bibinfo{person}{Kunlin Liu},
  \bibinfo{person}{Sugasa Marangonda}, \bibinfo{person}{Chris Umé},
  \bibinfo{person}{Mr. Dpfks}, \bibinfo{person}{Carl~Shift Facenheim},
  \bibinfo{person}{Luis RP}, \bibinfo{person}{Jian Jiang},
  \bibinfo{person}{Sheng Zhang}, \bibinfo{person}{Pingyu Wu},
  \bibinfo{person}{Bo Zhou}, {and} \bibinfo{person}{Weiming Zhang}.}
  \bibinfo{year}{2021}\natexlab{}.
\newblock \bibinfo{title}{DeepFaceLab: Integrated, flexible and extensible
  face-swapping framework}.
\newblock
\newblock
\urldef\tempurl%
\url{https://doi.org/10.1016/j.patcog.2023.109628}
\showDOI{\tempurl}
\showeprint[arxiv]{2005.05535}


\bibitem[Qian et~al\mbox{.}(2020)]%
        {DBLP:conf/eccv/QianYSCS20}
\bibfield{author}{\bibinfo{person}{Yuyang Qian}, \bibinfo{person}{Guojun Yin},
  \bibinfo{person}{Lu Sheng}, \bibinfo{person}{Zixuan Chen}, {and}
  \bibinfo{person}{Jing Shao}.} \bibinfo{year}{2020}\natexlab{}.
\newblock \showarticletitle{Thinking in Frequency: Face Forgery Detection by
  Mining Frequency-Aware Clues}. In \bibinfo{booktitle}{\emph{European
  Conference on Computer Vision}}. \bibinfo{publisher}{Springer},
  \bibinfo{pages}{86--103}.
\newblock
\urldef\tempurl%
\url{https://doi.org/10.1007/978-3-030-58610-2\_6}
\showDOI{\tempurl}


\bibitem[Ren et~al\mbox{.}(2021)]%
        {DBLP:conf/mm/RenXSYL21}
\bibfield{author}{\bibinfo{person}{Yifan Ren}, \bibinfo{person}{Xing Xu},
  \bibinfo{person}{Fumin Shen}, \bibinfo{person}{Yazhou Yao}, {and}
  \bibinfo{person}{Huimin Lu}.} \bibinfo{year}{2021}\natexlab{}.
\newblock \showarticletitle{{CAA:} Candidate-Aware Aggregation for Temporal
  Action Detection}. In \bibinfo{booktitle}{\emph{Proceedings of the 29th {ACM}
  international conference on Multimedia}}. \bibinfo{publisher}{{ACM}},
  \bibinfo{pages}{4930--4938}.
\newblock
\urldef\tempurl%
\url{https://doi.org/10.1145/3474085.3475616}
\showDOI{\tempurl}


\bibitem[R{\"{o}}ssler et~al\mbox{.}(2019)]%
        {DBLP:conf/iccv/RosslerCVRTN19}
\bibfield{author}{\bibinfo{person}{Andreas R{\"{o}}ssler},
  \bibinfo{person}{Davide Cozzolino}, \bibinfo{person}{Luisa Verdoliva},
  \bibinfo{person}{Christian Riess}, \bibinfo{person}{Justus Thies}, {and}
  \bibinfo{person}{Matthias Nie{\ss}ner}.} \bibinfo{year}{2019}\natexlab{}.
\newblock \showarticletitle{FaceForensics++: Learning to Detect Manipulated
  Facial Images}. In \bibinfo{booktitle}{\emph{IEEE International Conference on
  Computer Vision}}. \bibinfo{publisher}{{IEEE}}, \bibinfo{pages}{1--11}.
\newblock
\urldef\tempurl%
\url{https://doi.org/10.1109/ICCV.2019.00009}
\showDOI{\tempurl}


\bibitem[Song et~al\mbox{.}(2022)]%
        {DBLP:conf/mm/SongLFJCX22}
\bibfield{author}{\bibinfo{person}{Luchuan Song}, \bibinfo{person}{Xiaodan Li},
  \bibinfo{person}{Zheng Fang}, \bibinfo{person}{Zhenchao Jin},
  \bibinfo{person}{Yuefeng Chen}, {and} \bibinfo{person}{Chenliang Xu}.}
  \bibinfo{year}{2022}\natexlab{}.
\newblock \showarticletitle{Face Forgery Detection via Symmetric Transformer}.
  In \bibinfo{booktitle}{\emph{Proceedings of the 30th ACM International
  Conference on Multimedia}}. \bibinfo{publisher}{{ACM}},
  \bibinfo{pages}{4102--4111}.
\newblock
\urldef\tempurl%
\url{https://doi.org/10.1145/3503161.3547806}
\showDOI{\tempurl}


\bibitem[Tan et~al\mbox{.}(2021)]%
        {DBLP:conf/iccv/TanT0W21}
\bibfield{author}{\bibinfo{person}{Jing Tan}, \bibinfo{person}{Jiaqi Tang},
  \bibinfo{person}{Limin Wang}, {and} \bibinfo{person}{Gangshan Wu}.}
  \bibinfo{year}{2021}\natexlab{}.
\newblock \showarticletitle{Relaxed Transformer Decoders for Direct Action
  Proposal Generation}. In \bibinfo{booktitle}{\emph{IEEE International
  Conference on Computer Vision}}. \bibinfo{publisher}{{IEEE}},
  \bibinfo{pages}{13506--13515}.
\newblock
\urldef\tempurl%
\url{https://doi.org/10.1109/ICCV48922.2021.01327}
\showDOI{\tempurl}


\bibitem[Ulyanov et~al\mbox{.}(2017)]%
        {DBLP:journals/corr/UlyanovVL16}
\bibfield{author}{\bibinfo{person}{Dmitry Ulyanov}, \bibinfo{person}{Andrea
  Vedaldi}, {and} \bibinfo{person}{Victor Lempitsky}.}
  \bibinfo{year}{2017}\natexlab{}.
\newblock \bibinfo{title}{Instance Normalization: The Missing Ingredient for
  Fast Stylization}.
\newblock
\newblock
\showeprint[arxiv]{1607.08022}


\bibitem[Vahdani and Tian(2023)]%
        {DBLP:journals/pami/VahdaniT23}
\bibfield{author}{\bibinfo{person}{Elahe Vahdani} {and} \bibinfo{person}{Yingli
  Tian}.} \bibinfo{year}{2023}\natexlab{}.
\newblock \showarticletitle{Deep Learning-Based Action Detection in Untrimmed
  Videos: {A} Survey}.
\newblock \bibinfo{journal}{\emph{{IEEE} Trans. Pattern Anal. Mach. Intell.}}
  \bibinfo{volume}{45}, \bibinfo{number}{4} (\bibinfo{year}{2023}),
  \bibinfo{pages}{4302--4320}.
\newblock
\urldef\tempurl%
\url{https://doi.org/10.1109/TPAMI.2022.3193611}
\showDOI{\tempurl}


\bibitem[Vaswani et~al\mbox{.}(2017)]%
        {DBLP:conf/nips/VaswaniSPUJGKP17}
\bibfield{author}{\bibinfo{person}{Ashish Vaswani}, \bibinfo{person}{Noam
  Shazeer}, \bibinfo{person}{Niki Parmar}, \bibinfo{person}{Jakob Uszkoreit},
  \bibinfo{person}{Llion Jones}, \bibinfo{person}{Aidan~N Gomez},
  \bibinfo{person}{\L~ukasz Kaiser}, {and} \bibinfo{person}{Illia Polosukhin}.}
  \bibinfo{year}{2017}\natexlab{}.
\newblock \showarticletitle{Attention is All you Need}. In
  \bibinfo{booktitle}{\emph{Advances in Neural Information Processing
  Systems}}, Vol.~\bibinfo{volume}{30}. \bibinfo{publisher}{Curran Associates,
  Inc.}, \bibinfo{pages}{5998--6008}.
\newblock


\bibitem[Wang et~al\mbox{.}(2021)]%
        {DBLP:journals/pami/00010CJDZ0MTW0X21}
\bibfield{author}{\bibinfo{person}{Jingdong Wang}, \bibinfo{person}{Ke Sun},
  \bibinfo{person}{Tianheng Cheng}, \bibinfo{person}{Borui Jiang},
  \bibinfo{person}{Chaorui Deng}, \bibinfo{person}{Yang Zhao},
  \bibinfo{person}{Dong Liu}, \bibinfo{person}{Yadong Mu},
  \bibinfo{person}{Mingkui Tan}, \bibinfo{person}{Xinggang Wang},
  \bibinfo{person}{Wenyu Liu}, {and} \bibinfo{person}{Bin Xiao}.}
  \bibinfo{year}{2021}\natexlab{}.
\newblock \showarticletitle{Deep High-Resolution Representation Learning for
  Visual Recognition}.
\newblock \bibinfo{journal}{\emph{{IEEE} Trans. Pattern Anal. Mach. Intell.}}
  \bibinfo{volume}{43}, \bibinfo{number}{10} (\bibinfo{year}{2021}),
  \bibinfo{pages}{3349--3364}.
\newblock
\urldef\tempurl%
\url{https://doi.org/10.1109/TPAMI.2020.2983686}
\showDOI{\tempurl}


\bibitem[Wang et~al\mbox{.}(2016)]%
        {DBLP:conf/eccv/WangXW0LTG16}
\bibfield{author}{\bibinfo{person}{Limin Wang}, \bibinfo{person}{Yuanjun
  Xiong}, \bibinfo{person}{Zhe Wang}, \bibinfo{person}{Yu Qiao},
  \bibinfo{person}{Dahua Lin}, \bibinfo{person}{Xiaoou Tang}, {and}
  \bibinfo{person}{Luc~Van Gool}.} \bibinfo{year}{2016}\natexlab{}.
\newblock \showarticletitle{Temporal Segment Networks: Towards Good Practices
  for Deep Action Recognition}. In \bibinfo{booktitle}{\emph{European
  Conference on Computer Vision}}. \bibinfo{publisher}{Springer},
  \bibinfo{pages}{20--36}.
\newblock
\urldef\tempurl%
\url{https://doi.org/10.1007/978-3-319-46484-8\_2}
\showDOI{\tempurl}


\bibitem[Wang and Zhao(2022)]%
        {DBLP:conf/mm/WangZ22}
\bibfield{author}{\bibinfo{person}{Yongqi Wang} {and} \bibinfo{person}{Zhou
  Zhao}.} \bibinfo{year}{2022}\natexlab{}.
\newblock \showarticletitle{FastLTS: Non-Autoregressive End-to-End
  Unconstrained Lip-to-Speech Synthesis}. In
  \bibinfo{booktitle}{\emph{Proceedings of the 30th {ACM} International
  Conference on Multimedia}}. \bibinfo{publisher}{{ACM}},
  \bibinfo{pages}{5678--5687}.
\newblock
\urldef\tempurl%
\url{https://doi.org/10.1145/3503161.3548194}
\showDOI{\tempurl}


\bibitem[Xu et~al\mbox{.}(2018)]%
        {DBLP:conf/eccv/XuYFYYLPCH18}
\bibfield{author}{\bibinfo{person}{Ning Xu}, \bibinfo{person}{Linjie Yang},
  \bibinfo{person}{Yuchen Fan}, \bibinfo{person}{Jianchao Yang},
  \bibinfo{person}{Dingcheng Yue}, \bibinfo{person}{Yuchen Liang},
  \bibinfo{person}{Brian~L. Price}, \bibinfo{person}{Scott Cohen}, {and}
  \bibinfo{person}{Thomas~S. Huang}.} \bibinfo{year}{2018}\natexlab{}.
\newblock \showarticletitle{YouTube-VOS: Sequence-to-Sequence Video Object
  Segmentation}. In \bibinfo{booktitle}{\emph{European Conference on Computer
  Vision}}. \bibinfo{publisher}{Springer}, \bibinfo{pages}{603--619}.
\newblock
\urldef\tempurl%
\url{https://doi.org/10.1007/978-3-030-01228-1\_36}
\showDOI{\tempurl}


\bibitem[Yang et~al\mbox{.}(2023)]%
        {DBLP:journals/corr/abs-2209-00796}
\bibfield{author}{\bibinfo{person}{Ling Yang}, \bibinfo{person}{Zhilong Zhang},
  \bibinfo{person}{Yang Song}, \bibinfo{person}{Shenda Hong},
  \bibinfo{person}{Runsheng Xu}, \bibinfo{person}{Yue Zhao},
  \bibinfo{person}{Wentao Zhang}, \bibinfo{person}{Bin Cui}, {and}
  \bibinfo{person}{Ming-Hsuan Yang}.} \bibinfo{year}{2023}\natexlab{}.
\newblock \bibinfo{title}{Diffusion Models: A Comprehensive Survey of Methods
  and Applications}.
\newblock
\newblock
\showeprint[arxiv]{2209.00796}


\bibitem[Zeng et~al\mbox{.}(2020)]%
        {DBLP:conf/eccv/ZengFC20}
\bibfield{author}{\bibinfo{person}{Yanhong Zeng}, \bibinfo{person}{Jianlong
  Fu}, {and} \bibinfo{person}{Hongyang Chao}.} \bibinfo{year}{2020}\natexlab{}.
\newblock \showarticletitle{Learning Joint Spatial-Temporal Transformations for
  Video Inpainting}. In \bibinfo{booktitle}{\emph{European Conference on
  Computer Vision}}. \bibinfo{publisher}{Springer}, \bibinfo{pages}{528--543}.
\newblock
\urldef\tempurl%
\url{https://doi.org/10.1007/978-3-030-58517-4\_31}
\showDOI{\tempurl}


\bibitem[Zhang and Sim(2022)]%
        {DBLP:conf/icpr/ZhangS22}
\bibfield{author}{\bibinfo{person}{Bowen Zhang} {and} \bibinfo{person}{Terence
  Sim}.} \bibinfo{year}{2022}\natexlab{}.
\newblock \showarticletitle{Localizing Fake Segments in Speech}. In
  \bibinfo{booktitle}{\emph{26th International Conference on Pattern
  Recognition}}. \bibinfo{publisher}{{IEEE}}, \bibinfo{pages}{3224--3230}.
\newblock
\urldef\tempurl%
\url{https://doi.org/10.1109/ICPR56361.2022.9956134}
\showDOI{\tempurl}


\bibitem[Zhang et~al\mbox{.}(2022c)]%
        {DBLP:conf/eccv/ZhangWL22}
\bibfield{author}{\bibinfo{person}{Chen{-}Lin Zhang}, \bibinfo{person}{Jianxin
  Wu}, {and} \bibinfo{person}{Yin Li}.} \bibinfo{year}{2022}\natexlab{c}.
\newblock \showarticletitle{ActionFormer: Localizing Moments of Actions with
  Transformers}. In \bibinfo{booktitle}{\emph{European Conference on Computer
  Vision}}. \bibinfo{publisher}{Springer}, \bibinfo{pages}{492--510}.
\newblock
\urldef\tempurl%
\url{https://doi.org/10.1007/978-3-031-19772-7\_29}
\showDOI{\tempurl}


\bibitem[Zhang et~al\mbox{.}(2022b)]%
        {DBLP:conf/mm/ZhangLHW0G22}
\bibfield{author}{\bibinfo{person}{Daichi Zhang}, \bibinfo{person}{Fanzhao
  Lin}, \bibinfo{person}{Yingying Hua}, \bibinfo{person}{Pengju Wang},
  \bibinfo{person}{Dan Zeng}, {and} \bibinfo{person}{Shiming Ge}.}
  \bibinfo{year}{2022}\natexlab{b}.
\newblock \showarticletitle{Deepfake Video Detection with Spatiotemporal
  Dropout Transformer}. In \bibinfo{booktitle}{\emph{Proceedings of the 30th
  {ACM} International Conference on Multimedia}}. \bibinfo{publisher}{{ACM}},
  \bibinfo{pages}{5833--5841}.
\newblock
\urldef\tempurl%
\url{https://doi.org/10.1145/3503161.3547913}
\showDOI{\tempurl}


\bibitem[Zhang et~al\mbox{.}(2022a)]%
        {DBLP:conf/eccv/ZhangFL22}
\bibfield{author}{\bibinfo{person}{Kaidong Zhang}, \bibinfo{person}{Jingjing
  Fu}, {and} \bibinfo{person}{Dong Liu}.} \bibinfo{year}{2022}\natexlab{a}.
\newblock \showarticletitle{Flow-Guided Transformer for Video Inpainting}. In
  \bibinfo{booktitle}{\emph{European Conference on Computer Vision}}.
  \bibinfo{publisher}{Springer}, \bibinfo{pages}{74--90}.
\newblock
\urldef\tempurl%
\url{https://doi.org/10.1007/978-3-031-19797-0\_5}
\showDOI{\tempurl}


\bibitem[Zhang et~al\mbox{.}(2023)]%
        {DBLP:journals/tkde/ZhangCWP23}
\bibfield{author}{\bibinfo{person}{Yuxin Zhang}, \bibinfo{person}{Yiqiang
  Chen}, \bibinfo{person}{Jindong Wang}, {and} \bibinfo{person}{Zhiwen Pan}.}
  \bibinfo{year}{2023}\natexlab{}.
\newblock \showarticletitle{Unsupervised Deep Anomaly Detection for
  Multi-Sensor Time-Series Signals}.
\newblock \bibinfo{journal}{\emph{{IEEE} Trans. Knowl. Data Eng.}}
  \bibinfo{volume}{35}, \bibinfo{number}{2} (\bibinfo{year}{2023}),
  \bibinfo{pages}{2118--2132}.
\newblock
\urldef\tempurl%
\url{https://doi.org/10.1109/TKDE.2021.3102110}
\showDOI{\tempurl}


\bibitem[Zhao et~al\mbox{.}(2021)]%
        {DBLP:conf/cvpr/ZhaoZ0WZY21}
\bibfield{author}{\bibinfo{person}{Hanqing Zhao}, \bibinfo{person}{Wenbo Zhou},
  \bibinfo{person}{Dongdong Chen}, \bibinfo{person}{Tianyi Wei},
  \bibinfo{person}{Weiming Zhang}, {and} \bibinfo{person}{Nenghai Yu}.}
  \bibinfo{year}{2021}\natexlab{}.
\newblock \showarticletitle{Multi-Attentional Deepfake Detection}. In
  \bibinfo{booktitle}{\emph{Proceedings of the IEEE Conference on Computer
  Vision and Pattern Recognition}}. \bibinfo{publisher}{{IEEE}},
  \bibinfo{pages}{2185--2194}.
\newblock
\urldef\tempurl%
\url{https://doi.org/10.1109/CVPR46437.2021.00222}
\showDOI{\tempurl}


\bibitem[Zhou et~al\mbox{.}(2017)]%
        {DBLP:conf/cvpr/ZhouHMD17}
\bibfield{author}{\bibinfo{person}{Peng Zhou}, \bibinfo{person}{Xintong Han},
  \bibinfo{person}{Vlad~I. Morariu}, {and} \bibinfo{person}{Larry~S. Davis}.}
  \bibinfo{year}{2017}\natexlab{}.
\newblock \showarticletitle{Two-Stream Neural Networks for Tampered Face
  Detection}. In \bibinfo{booktitle}{\emph{Proceedings of the IEEE Conference
  on Computer Vision and Pattern Recognition Workshops}}.
  \bibinfo{publisher}{{IEEE}}, \bibinfo{pages}{1831--1839}.
\newblock
\urldef\tempurl%
\url{https://doi.org/10.1109/CVPRW.2017.229}
\showDOI{\tempurl}


\bibitem[Zhou and Lim(2021)]%
        {DBLP:conf/iccv/ZhouL21}
\bibfield{author}{\bibinfo{person}{Yipin Zhou} {and} \bibinfo{person}{Ser{-}Nam
  Lim}.} \bibinfo{year}{2021}\natexlab{}.
\newblock \showarticletitle{Joint Audio-Visual Deepfake Detection}. In
  \bibinfo{booktitle}{\emph{{IEEE} International Conference on Computer
  Vision}}. \bibinfo{publisher}{{IEEE}}, \bibinfo{pages}{14780--14789}.
\newblock
\urldef\tempurl%
\url{https://doi.org/10.1109/ICCV48922.2021.01453}
\showDOI{\tempurl}


\bibitem[Zi et~al\mbox{.}(2020)]%
        {DBLP:conf/mm/ZiCCMJ20}
\bibfield{author}{\bibinfo{person}{Bojia Zi}, \bibinfo{person}{Minghao Chang},
  \bibinfo{person}{Jingjing Chen}, \bibinfo{person}{Xingjun Ma}, {and}
  \bibinfo{person}{Yu{-}Gang Jiang}.} \bibinfo{year}{2020}\natexlab{}.
\newblock \showarticletitle{WildDeepfake: {A} Challenging Real-World Dataset
  for Deepfake Detection}. In \bibinfo{booktitle}{\emph{Proceedings of the 28th
  {ACM} International Conference on Multimedia}}. \bibinfo{publisher}{{ACM}},
  \bibinfo{pages}{2382--2390}.
\newblock
\urldef\tempurl%
\url{https://doi.org/10.1145/3394171.3413769}
\showDOI{\tempurl}


\end{thebibliography}

\clearpage
\appendix
\section{Appendix}
\subsection{Comparison between Existing Audio and Video Forensics Datasets}
\label{app:a1}
We present a comprehensive analysis of the recently popular datasets for audio and video forensics. Table~\ref{tab:comparison-dataset} summarizes the benchmark datasets that have been used for research on detecting generative content of audio and video, particularly in deepfake detection. While most of the existing datasets~\cite{DBLP:conf/mm/ZiCCMJ20} focused on simple binary classification tasks related to facial image manipulation, recent advancements in deepfake detection technology resulted in the emergence of binary classification tasks for audio~\cite{DBLP:conf/nips/FrankS21}  and multimodal audio-visual data~\cite{DBLP:conf/nips/KhalidTKW21}. These datasets involve not only facial but also audio information, and their emergence is indicative of the rapid development of deepfake detection technology. Lav-DF~\cite{DBLP:journals/corr/abs-2204-06228} and Psynd~\cite{DBLP:conf/icpr/ZhangS22}, two other emerging datasets for TFL, are manipulated based on semantic content, making the attacks on these types of videos more similar to real-world scenarios. However, research on these datasets has remained limited to scenes related to human faces and speeches, which is only a small part of the AIGC task. Furthermore, while there are emerging classification datasets~\cite{DBLP:conf/wacv/MittalSSCM23} for tampering detection beyond face, the quantity of these datasets is limited due to manual generation. To expand the scope of research, it is important to develop more diverse datasets for TFL tasks. Our datasets, similar to TFL subset of ForgeryNet~\cite{DBLP:conf/cvpr/HeGCZYSSS021} , uses a random approach to generate segments, which facilitates research on TFL tasks beyond facial images and beyond binary classification tasks. Moreover, our generation process can be reproduced under low-cost conditions. Based on our research, the dataset can further expand to more diverse scenarios and promote TFL task research.

\begin{table}[htbp]
\centering
\caption{Quantitative comparison of TVIL to existing popular Video and Audio Forensics Datasets in recent 3 years. Cls: Classification; SL: Spatial Localization; TFL: Temporal Forgery Localization; V: Visual; A: Audio.}
\label{tab:comparison-dataset}
\resizebox{\columnwidth}{!}{%
\begin{tabular}{ccccccccc}
\hline
Dataset           & Year & Tasks      & Modality & Application        & Manipulated         & \# Attacks & \#Real & \#Fake  \\ \hline
WildDeepfake~\cite{DBLP:conf/mm/ZiCCMJ20}      & 2021 & Cls        & V        & Face               & AIGC                & -          & 3,805  & 3,509   \\
FakeAVCeleb~\cite{DBLP:conf/nips/KhalidTKW21}       & 2021 & Cls        & A+V      & Face               & AIGC                & 3          & 570    & 19,500  \\
ForgeryNet~\cite{DBLP:conf/cvpr/HeGCZYSSS021}        & 2021 & SL/TFL/Cls & V        & Face               & AIGC                & 5          & 99,630 & 121,617 \\
Lav-DF~\cite{DBLP:journals/corr/abs-2204-06228}            & 2022 & TFL/Cls    & A+V      & Face               & AIGC                & 2          & 36,431 & 99,873  \\ \hline
WaveFake~\cite{DBLP:conf/nips/FrankS21}          & 2021 & Cls        & A        & Speech             & AIGC                & 6          & 18,100 & 104,885 \\
Psynd~\cite{DBLP:conf/icpr/ZhangS22}             & 2022 & TFL        & A        & Speech             & AIGC                & 1          & 30     & 2371    \\ \hline
VideoSham~\cite{DBLP:conf/wacv/MittalSSCM23}         & 2023 & Cls        & A+V      & Video Manipulation & User Generated      & 40         & 413    & 413     \\
TVIL(Ours)        & 2023 & TFL        & V        & Video Manipulation & AIGC                & 4          & 914    & 3539    \\ \hline
\end{tabular}%
}
\end{table}

\subsection{More Experiments Results for LAV-DF Subset}
\label{appendix:lavdf-subset}
 
We present the AP and AR performance of our method and state-of-the-art algorithms on the Lav-DF subset in Table~\ref{tab:lavdf_subset}. This subset exclusively contains manipulated videos with visual forgeries, excluding those with audio-only modifications. The results show that the single-modal algorithms in Table~\ref{tab:lavdf_subset} outperform their counterparts in Table~\ref{tab:tvil_full}, validating the efficacy of using visual information alone in this context. Notably, our method achieves state-of-the-art performance with an AP@0.5 of $98.83\%$ using solely visual modality.

Despite our method's adaptability to different modalities, applying techniques like contrastive learning to analyze modal inconsistencies during multi-modal fusion posed challenges. Introducing the audio modality in this subset, where authenticity is independent of audio, might lead to a decrease in overall performance. Feature fusion could potentially confuse critical features within this specific subset, as evident from the performance drop of the multi-modal algorithm AVFusion~\cite{DBLP:conf/visapp/BagchiMFS22}.

Nevertheless, our proposed method consistently outperforms other algorithms for both uni- and multi-modal inputs. Moreover, by incorporating audio features, our model achieves a substantial $7.41\%$ improvement in AP at tIoU 0.95, demonstrating the robustness and adaptability of our approach in multi-modal scenarios.

\begin{table}[htbp]
\centering
\caption{Performance comparison on Lav-DF Sub Set. Bold faces correspond to the top performance.}
\label{tab:lavdf_subset}
\resizebox{\columnwidth}{!}{%
\begin{tabular}{cc|ccccccc}
\hline
\multirow{2}{*}{Methods} & \multirow{2}{*}{Feature} & \multicolumn{7}{c}{Sub Set}                                                                                          \\ \cline{3-9} 
                         &                          & AP@0.5         & AP@0.75        & AP@0.95        & AR@10          & AR@20          & AR@50          & AR@100         \\ \hline
MDS~\cite{DBLP:conf/mm/ChughGDS20}                      & Visual                   & 23.43          & 3.48           & 00.00          & 58.53          & 56.68          & 53.16          & 49.67          \\
AGT~\cite{DBLP:journals/corr/abs-2101-08540}                      & Visual                   & 15.69          & 10.69          & 00.15          & 49.11          & 40.31          & 31.70          & 23.13          \\
BMN~\cite{DBLP:conf/iccv/LinLLDW19}                      & Visual                   & 32.32          & 11.38          & 00.14          & 59.69          & 48.17          & 39.01          & 34.17          \\
BMN (I3D)~\cite{DBLP:conf/iccv/LinLLDW19}               & Visual                   & 28.10          & 5.47           & 00.01          & 55.49          & 54.44          & 52.14          & 47.72          \\
AVFusion~\cite{DBLP:conf/visapp/BagchiMFS22}                & Visual+Audio             & 62.01          & 22.77          & 00.11          & 61.98          & 58.08          & 53.31          & 50.52          \\
\multirow{2}{*}{BA-TFD~\cite{DBLP:journals/corr/abs-2204-06228}}  & Visual                   & 83.55          & 41.88          & 00.24          & 65.79          & 62.30          & 57.95          & 55.34          \\
                         & Visual+Audio             & 85.20          & 47.06          & 00.29          & 67.34          & 64.52          & 61.19          & 59.32          \\
ActionFormer~\cite{DBLP:conf/eccv/ZhangWL22}            & Visual                   & 98.06          & 94.43          & 27.25          & 91.30          & 92.04          & 92.27          & 92.28          \\ \hline
\multirow{2}{*}{Ours}    & Visual                   & \textbf{98.83} & \textbf{95.95} & 30.11          & \textbf{92.32} & \textbf{92.65} & \textbf{92.74} & \textbf{92.75} \\
                         & Visual+Audio             & 98.54          & 94.30          & \textbf{37.52} & 91.61          & 91.97          & 92.06          & 92.06          \\ \hline
\end{tabular}%
}
\end{table}

\subsection{More Experiments Results for Video-level Face Forgery Classification}
 \label{appendix:lavdf-classification}

We also conducted a comparison between our method and previous deepfake detection methods on the Lav-DF Full Set for the video-level forgery classification task. The evaluation metric used is the Area Under the Receiver Operating Characteristic Curve (AUC), and the results are summarized in Table~\ref{tab:classification_result}. In our approach, we utilize the scores obtained from detected forgery timestamps as the classification scores for the respective videos. As observed, frame-based algorithms such as $F^3$-Net~\cite{DBLP:conf/eccv/QianYSCS20} exhibit significant performance degradation in classifying partially manipulated videos due to their lack of consideration of temporal factors, leading to substantial discrepancies in discriminating between different frames. On the other hand, video-level algorithms such as EfficientViT~\cite{DBLP:conf/iciap/CoccominiMGF22} demonstrated relatively effective recognition of deepfake videos with partially manipulated segments, but are unable to provide corresponding timestamps for the forgeries. In contrast, our method achieved the best classification performance while also providing corresponding forgery timestamps. Additionally, our temporal forgery localization performance significantly outperforms MDS~\cite{DBLP:conf/mm/ChughGDS20} and BA-TFD~\cite{DBLP:journals/corr/abs-2204-06228}. It is worth noting that our model was not specifically designed for the classification task, and further performance improvement could be achieved by introducing a dedicated classification head.

\begin{table}[htbp]
\centering
\caption{Deepfake detection results on the Lav-DF dataset. Bold
faces correspond to the top performance}
\label{tab:classification_result}
\resizebox{0.4\columnwidth}{!}{%
\begin{tabular}{cc}
\hline
Methods                                            & AUC  \\ \hline
$F^3$-Net~\cite{DBLP:conf/eccv/QianYSCS20}        & 52.0   \\
MDS~~\cite{DBLP:conf/mm/ChughGDS20}                & 82.8 \\
EfficientViT~\cite{DBLP:conf/iciap/CoccominiMGF22} & 96.5 \\
BA-TFD~\cite{DBLP:journals/corr/abs-2204-06228}    & 99.0 \\ \hline
Ours                                               & \textbf{99.8} \\ \hline
\end{tabular}%
}
\end{table}

\end{document}